\documentclass[aps,prc,twocolumn,showpacs]{revtex4}
\usepackage{epsfig,amsmath,amssymb}

\newcommand{\be}{\begin{equation}}
\newcommand{\ee}{\end{equation}}
\newcommand{\bea}{\begin{eqnarray}}
\newcommand{\eea}{\end{eqnarray}}
\newcommand{\fp}{f_\Psi}

\begin{document}



\title{ Kinetic description of charmonium production
 in high-energy nuclear collisions}

\author{Alberto Polleri$^{\,a,b}$}
\author{Thorsten Renk$^{\,a}$}
\author{Roland Schneider$^{\,a}$}
\author{Wolfram Weise$^{\,a,b}$}


\affiliation{ $^{a}$Physik Department, Technische Universit\"{a}t M\"{u}nchen,
D-85747 Garching, Germany\\
 $^{b}$ECT$^\ast$, Villa Tambosi, I-38050 Villazzano (Trento), Italy}

\begin{abstract}
We study the evolution of charmonia as they collide with the constituents 
of the fireball produced in  high-energy nucleus-nucleus collisions. The 
latter evolves in a manner controlled by the equation of state as given 
by lattice QCD, and is constructed in such a way that the observed hadronic
spectra are correctly reproduced. A kinetic description of charmonium 
interactions with both quark-gluon and hadronic degrees of freedom allows 
to study in  detail the evolution in different regimes, controlled by 
collision energy, kinematics and geometry. The data collected at the CERN-SPS
accelerator are well described and new estimates for $J/\psi$ production at
BNL-RHIC are presented.
\end{abstract}


\maketitle

\section{Introduction}
\label{sec_introduction}

The very first suggestion to study $J/\psi$ production and suppression
in heavy-ion
collisions \cite{MS86} was based on a two-fold observation: First, the
energy density accumulated in such collisions is expected to be
high enough that the medium formed in the collision presumably undergoes 
a transition from
hadronic matter to a quark-gluon plasma (QGP). Second, $J/\psi$ is
bound essentially by the confinement force which, when screened in the
QGP, loses its binding character. Produced $J/\psi$s dissolve and their
charm quarks end up in pairs of $D$ mesons. The expectation is therefore
that the  $J/\psi$ production cross section is suppressed in a
hot and dense environment as compared with extrapolations from proton-proton
($pp$) collisions.

Indeed, lattice simulations \cite{lat1} indicate that  QCD with (2+1) flavors
undergoes a
transition at a temperature of about 170  MeV \cite{lat2} from a
confined hadronic phase to a phase where quarks and gluons constitute
the relevant active degrees of freedom. The possible existence of this new
state of matter has raised enormous interest, and the search for
experimental evidence of its creation continues since about 20
years. Yet there is still no unambiguous proof that the QGP has
been produced in the laboratory. A primary reason for this is
that any QGP signature, including $J/\psi$ suppression, is folded with
the time evolution of the fireball created in such a
collision. Moreover, the evolution continues after the supposedly
produced plasma undergoes a transition from partonic to hadronic
degrees of freedom. Any information on the early stage is then hidden
behind signals from the hadronic phase. Finally, the most  important
uncertainty is perhaps the poor knowledge of the process of
thermalization.  The particularly interesting case of $J/\psi$
suppression is thus subject  of a vigorous experimental search and
theoretical debate. The first systematic signs of suppression
\cite{NA38} were later on explained by more conventional mechanisms
\cite{GeHu}, already present in proton-nucleus ($pA$) reactions. Only
with the beginning of the experimental search using $Pb\!+\!Pb$ collisions
the first signals of anomalous suppression, beyond extrapolations from
$pA$ collisions, were seen \cite{NA50one}.

While much phenomenology has been developed in order to explain the
latest SPS data \cite{NA50two}, it is still debated whether or not
anomalous suppression can be attributed to QGP formation. In fact,
realistic models can be constructed which incorporate well known nuclear
effects such as initial state gluon radiation \cite{HK98}, color
excitation \cite{HKP01}, initial state parton energy loss \cite{KN84}
and coherence effects \cite{KTH01}. All of these can potentially
account for the observed data because they provide the necessary
non-linear dependence on the number of participating particles which
becomes significant only when going beyond $pA$ collisions.

In view of the uncertainties inherent in the construction of models
which consider such  effects, we attempt an approach to the problem
which maximizes the number of  independent physical constraints. The
usual strategy is to attribute the missing suppression to medium
effects, exclusively designed to handle the  $J/\psi$ case alone.  For
this procedure to be conclusive one needs a very reliable baseline for
the  description of the production process. Here we adopt a different
and novel perspective and make use of the knowledge of the medium
evolution as inferred by a variety of other  observables. We then
explore whether the same description is  consistent with the observed
$J/\psi$ measurements. We stress that this  is possible because the
specific time evolution of the medium is not constructed  or fitted in
order to reproduce the  $J/\psi$ suppression effect, but it is
constrained beforehand by other independent data.

We treat charmonium production in nucleus-nucleus ($AB$)
collisions  as a two-step process, factorizing direct production from
the subsequent evolution in the medium. The first part includes the
conventional description \cite{KLNS96}  of nuclear effects within the
Glauber model framework, with values of the absorption cross section
either extrapolated from the
suppression observed experimentally already in $pA$ collisions, or
fitted to reproduce the results of more sophisticated
computations \cite{KTH01,C02}.  The second part is
a description within kinetic theory of the interaction of $J/\psi$s
with the different degrees of freedom that populate the evolving
medium at different times.  In this framework interactions are
realized by cross sections, while the evolution of the medium  itself
is constrained by an independent description of hadronic
spectra. With this procedure we are able to
specify the nature of the medium and to follow  the detailed time
evolution of $J/\psi$ mesons.

It might seem natural at this stage to also incorporate the mentioned
Debye screening of the inter-quark potential, for example by assuming
complete $J/\psi$ dissolution above a certain temperature
\cite{DPS01,W01}.  However, it is not clear what an all-embracing
description of the  interaction of the $J/\psi$ with the medium should
be. Screening arises from the interaction of the virtual
gluons that bind  the $J/\psi$ with the gluons of the
environment. Such a process can also be viewed as
scattering of the $J/\psi$, that fluctuates into a $c\bar{c}$ pair and
non-perturbative multi-gluon exchanges, off the thermalized
gluons \cite{HLZ88}. Taking into account both screening and kinetic 
scattering can
amount to a certain degree to double counting.  Disentangling these
effects even qualitatively, however, is a highly non-trivial task
which requires a dedicated analysis.  At this stage, we can say that
screening, as inferred from lattice QCD, is a purely  static
concept. Since $J/\psi$s are not produced at rest with
respect  to the medium, a kinetic approach seems to be appropriate
within the present  context. Obviously, a more rigorous treatment of
this question is required in the future.

The paper is structured in four sections. First, the initial
conditions for  the solution of the kinetic equation are discussed. We
provide a simple extension of the parton model prescription for the
$pp \rightarrow c \bar c\, X$ process, scaling the open charm
production cross section with the average number of  collisions. At
the same time charmonium production is computed treating
nuclear suppression as effective Glauber absorption. We take into account
the role of excited charmonium states $\psi'$ and $\chi$ which
contribute sequentially to $J/\psi$ production. We will
often denote charmonia generically by $\Psi$ as a shorthand notation.

In the following section a thorough description of the produced medium
is given. This medium is assumed to thermalize and then described
within a  thermodynamically consistent quasi-particle picture
\cite{SW01}, incorporating important  aspects of confinement. This
approach treats quarks and gluons as massive thermal quasi-particles,
with their properties determined to reproduce lattice QCD results. The
driving force of the transition, the confinement/deconfinement
process, is given a statistical meaning in terms of a reduction of the
thermally active partonic degrees of freedom as the critical
temperature is approached  from above.  The fireball evolution is
fixed by requiring  agreement with hadronic observables at freeze-out
\cite{RSW02}.  Thermodynamics is calculated along volumes of constant
proper time under the assumption of total entropy conservation and
using the equation of state (EoS) from lattice QCD as described by the 
quasi-particle model. In a self-consistent
calculational procedure, we determine temperature and  volume of the
system as a function of proper time.

The main part of the paper is then focused on the kinetic description
of charmonium evolution in the fireball. A transport equation is
presented, including dissociation processes in the collision term. In
the QGP, $\Psi$ dissociation is provided by the well known
Bhanot-Peskin cross section  \cite{P79,BP79}. We also take into
account the possibility of $\Psi$ regeneration in the QGP. As recently
suggested by several authors  \cite{BMS00,TSR01,GR01,GOR01}, this
process  can potentially overwhelm dissociation, leading to $\Psi$
enhancement once the heavy ion collision energy is large enough. The
coalescence process of $c \bar c$ quark pairs into $\Psi$ is
controlled by the cross section obtained from that of dissociation
using detailed balance. A simplified, approximate kinetic equation is
then obtained. Its solution can be written in closed form, with time
dependence entering through the evolution of fireball  temperature and
volume.

Finally, results for $\Psi$ production at SPS and RHIC are presented
and a thorough discussion of theoretical uncertainties concludes the
paper. 

\section{Charm Production off Nuclei: Initial Conditions}
\label{sec_incond}

\subsection{Open charm}

The production of charmed quarks is commonly described within
perturbative QCD. The perturbative approach is strictly  valid only
for processes involving large gluon virtualities, usually provided by
high-momentum exchange or large masses. In the present case the
smallest scale is that of the charm quark mass $m_c \simeq 1.5$ GeV,
implying that perturbation theory at lowest order receives important
corrections.  Moreover, computations are accurate only at large enough
transverse momenta $p_T$, whereas small values of $p_T$ contribute
significantly to total yields.

In the following discussion we restrict ourselves to the leading
order treatment presented in \cite{ELL89,GAV94}, with suitable
adjustments in order to meet phenomenology. We consider the leading
processes $q \bar q \rightarrow c \bar c$ and $g g \rightarrow c \bar
c$.  In terms of their elementary differential cross sections $d{\hat
\sigma}_i/d{\hat t}$, the spectrum of $c$ quarks produced in $pp$
collisions at rapidity $y_c$ and transverse momentum $p_T$ is 
\be
\frac{d\sigma_{pp \rightarrow c}}{dy_c\,dp^2_T} \!=\! K\!\! \int\! dy_{\bar c}
\!\!\!\!\!  \sum_{i = u,d,s,g}\!\!\!\! F_{i/p}(x_1,\mu_c^2)\,
F_{i/p}(x_2,\mu_c^2) \, \frac{d{\hat \sigma}_i}{d{\hat t}},
\label{opencharm}
\ee 
where $x_{1,2} = (m^c_T/\sqrt{s})\,[\exp(\pm y_{\bar c}) + \exp(\pm y_c)]$ 
are the momentum fractions of the partons in the
colliding protons and $F_{i/p} = x_i f_{i/p}$. The rapidity variable is
defined as usual as $y = 1/2\, \log[\,(E+p_z)/(E-p_z)\,]$. The factorization
scale is taken at $\mu_c = 1.4$ GeV, which is of the order of the $c$
quark mass, and a scaling factor $K = 2$ is used.  We
employ the GRV94LO \cite{GRV94} parton distributions $f_{i/p}$ and neglect 
the effect of intrinsic transverse parton momentum. Moreover we set in
eq.~(\ref{opencharm}), where appropriate, $m_c = \mu_c$. In this way it is
possible to effectively reproduce to good accuracy the
next-to-leading order computations for total ($y_c$- and
$p_T$-integrated) charm production as given in \cite{GAV94} and
conveniently parametrized in \cite{GOR01} as
\be
\sigma_{pp \rightarrow c}^{NLO}(s) = \sigma_c^0 (1 - m_0/\sqrt{s})^{8.185}
(\sqrt{s}/m_0)^{1.132},
\ee
where $\sigma_c^0 = 3.392\ \mu$b and $m_0 = 2.984$ GeV.

To estimate the spectrum of $c$ quarks in $AB$ collisions, the simplest
approach is to scale the $pp$ result with the overlap function 
\be
T_{AB}(b) = \int\! d^2 r\ \,T_A(r)\, T_B(\tilde r)\,
\ee
where $T_{A,B}(b) = \int dz\, \rho_{A,B}(z,b)$ are the usual overlap
functions, expressed in terms of longitudinal integrals over the nuclear
densities $\rho_{A,B}$, and $\tilde r = |\vec b - \vec r|$, being $\vec r$
the transverse coordinate and $\vec b$ the impact parameter.

\begin{figure}[t]
\begin{center}
\epsfig{file=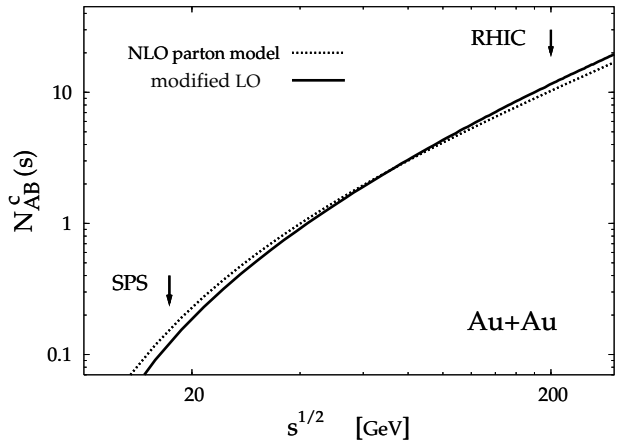,width=6cm,angle=-90}
\caption{Number of $c \bar c$ pairs produced in a $Au+Au$ collision as
function of the center of mass energy, computed by integrating the
spectrum given in eq.~(\ref{charmAB}).  Curves compare the NLO parton
model result (dotted curve) with modified LO  calculations with $\mu_c = 1.4$
GeV (solid).  From SPS to
RHIC energy the number of produced pairs grows by nearly two orders
of magnitude}
\label{figure:numcharm}
\end{center}
\end{figure}

On the other hand, as the collision energy increases, shadowing
effects are expected to become important, reducing the total yield \cite{KT02}.
Another correction to be introduced is the broadening effect on the colliding
partons' transverse momenta, resulting in a broader $p_T$ spectrum of 
produced charmed quarks. For simplicity we neglect these effects, thereby 
obtaining an upper limit for the $c$ quark production cross section.
The charmed quark spectrum in AB collisions is then computed as
\be
\frac{dN^c_{AB}}{dy_c\,dp^2_T}(b) = 
\frac{d\sigma_{pp \rightarrow c}}{dy_c\,dp^2_T} 
\ T_{AB}(b)\,.
\label{charmAB}
\ee
The total number of charm quarks produced, for example, in a $Au\!+\!Au$ 
collision can
be obtained by integrating the spectrum given in the latter
equation. The result is shown in Fig.~(\ref{figure:numcharm}). One 
observes a great difference between the values at SPS and RHIC energies, 
amounting to nearly two orders of magnitude.
Moreover, note that
\be
N^c_{AB}(s) = \sigma_{pp \rightarrow c}(s)\ T_{AB}(b) =
N^c_{pp}(s)\ N^{coll}_{AB}(b)\,,
\ee
where $N^c_{pp} = \sigma_{pp \rightarrow c} / \sigma^{in}_{pp}$ and
$N^{coll}_{AB}(b) = \sigma^{in}_{pp}\,T_{AB}$, being 
$\sigma^{in}_{pp}$ the inelastic $pp$ cross section. Since for large nuclei 
$N^{coll}_{AB}(b=0) > 1000$, it is clear how many more $c$ quarks are 
produced in a $AB$ collision with respect to the $pp$ case.

\subsection{Hidden Charm}

The description of charmonium production in nuclear collisions is a
more complicated task as compared to open $c \bar c$ production. At the root
of the problem lies the fact that exclusive production of composite particles 
in hadronic collisions is basically a non-perturbative process.
Only at large transverse momenta it is possible to make robust predictions 
for the spectra. 
Nevertheless, at least at the phenomenological level, much work has
been done in order to understand the physics underlying the results
of several experiments. Let us examine $pp$ collisions

We consider first of all the total production cross section and make use 
of the parametrization given in \cite{S94}. Since it was obtained by fitting
low-energy data ($\sqrt{s} < 50$ GeV), we modified it to simulate the 
high energy behavior in terms of NLO effects in the color evaporation 
model \cite{VOGTREP}. We parametrize
\be
\sigma_{pp \rightarrow J/\psi}(s) = \sigma_{J/\psi}^0\, 
(1 - m_{J/\psi}/\sqrt{s})^9\,
(1 + 0.6\,\sqrt{s}/m_{J/\psi})\,,
\ee
with $\sigma_{J/\psi}^0 =$ 100 nb and $m_{J/\psi} = 3.1$ GeV. The above 
formula gives essentially identical results as that of \cite{S94} in 
the low energy region and is in line with the recent PHENIX measurement
at RHIC \cite{PHENIX}.
The rapidity modulation
can be inferred from the relation $d\sigma/dy \sim x_1g(x_1)\,x_2g(x_2)$
where $g(x) \sim (1-x)^5/x$ is taken to be the gluon distribution in the 
proton and $x_{1,2} = (m_\Psi/\sqrt{s})\,\exp(\pm y)$.
We can then write the transverse-momentum-integrated $\Psi$ production 
spectrum in $pp$ collisions as
\be
\frac{d\sigma_{pp \rightarrow {J/\psi}}}{dy} = \sigma_{pp \rightarrow 
{J/\psi}}(s)\ F(s,y)\,,
\label{hiddencharm}
\ee
where the $y$-dependent part is
\bea
F(s,y) \!&=&\! C(s)\ (1 - x_1)^5\,(1 - x_2)^5 \nonumber \\
&=&\! C(s)\,\left[ 1-2(m_{J/\psi}/\sqrt{s})\cosh y + m^2_{J/\psi}/s \right]^5
\eea
while $C(s)$ is chosen to satisfy the normalization constraint 
$\int\! dy\, F(s,y) = 1$. The transverse momentum dependence will not
be needed in the following, except for the value of $\langle p_T^2 \rangle$
which can be taken directly from experiment.

We now consider nuclear effects, starting with the simpler case of 
proton-nucleus ($pA$) collisions. It has been shown that the experimental 
results on ${J/\psi}$ production can be described using
\be
\sigma_{pA \rightarrow {J/\psi}} = \sigma_{pp \rightarrow {J/\psi}} \int\! d^2b
\ T_A(b)\ S^{abs}_A(b)\,
\label{protnuc}
\ee
for the total production cross section. The factor
\be
S^{abs}_A(b) =
\frac{1 - \exp\left[-\sigma^{abs}_{{J/\psi} N}\,T_A(b)\right]}
{\sigma^{abs}_{{J/\psi} N}\,T_A(b)}
\label{nucsupp}
\ee
is the survival probability for ${J/\psi}$ to escape the nucleus without
being  dissociated. It includes the effective absorption cross section
$\sigma^{abs}_{{J/\psi} N}$, a quantity of the order of $3$ mb for
mid-rapidity ${J/\psi}$s as measured at $E_{lab} = 800$ GeV at  Fermilab,
while it amounts to $4-6$ mb for mid-rapidity ${J/\psi}$s as  measured at
$E_{lab} = 200-400$ GeV at the SPS. The absorption cross  section
parametrizes various poorly known effects, with varying importance
depending on the collision energy. Among those effects are the
presence of color non-singlet degrees of freedom in the dynamics of colliding
nucleons, initial state parton energy loss and coherence length and
shadowing effects. Moreover, $\sigma^{abs}_{J/\psi N}$ contains information
on the nuclear absorption of the excited states $\psi'$ and $\chi_c$ which 
feed into the observed ${J/\psi}$ yield.

A common property of all of the above effects is the approximate
linear dependence on the length of the path which ${J/\psi}$ traverses in the
nuclei. It is therefore justified, to first approximation, 
to use the same ``optical'' absorption
formula eq.~(\ref{nucsupp}), provided a re-scaling and re-interpretation 
of  $\sigma^{abs}_{J/\psi N} \rightarrow \sigma^{NUC}_{J/\psi N}$ and 
$S^{abs}_{A,B} \rightarrow S^{NUC}_{A,B}$ is done.

When looking at ${J/\psi}$ production in $AB$ collisions,
one can estimate the cross section for a given impact parameter by 
generalizing eq.~(\ref{protnuc}). The effects 
of the produced medium are the central topic of this paper and will
be thoroughly discussed in the following sections. Neglecting them for the
moment, one obtains
\be
\frac{dN^{J/\psi}_{AB}}{dy}(b) = \frac{d\sigma_{pp \rightarrow {J/\psi}}}{dy}
\ T_{AB}(b)\ S^{NUC}_{AB}(b)\,,
\label{initialpsi}
\ee
where all nuclear effects are included in the suppression function
\be
S^{NUC}_{AB}(b) \!=\! T_{AB}^{-1}(b)\! \int\! d^2 r
\ T_A(r)\, S^{NUC}_A(r)\ T_B(\tilde r)\, S^{NUC}_B(\tilde r)
\ee
which has the obvious property $S^{NUC}_{AB} < 1$ and 
$S^{NUC}_{AB} \rightarrow 1$ if $\sigma^{NUC}_{{J/\psi} N} \rightarrow 0$.

Since nuclear effects depend on energy, we have chosen 
$\sigma^{NUC}_{{J/\psi} N}(s_0) = 4.5$ mb at the SPS energy 
$\sqrt{s_0} = 17.3$ GeV ($E_{lab} = 158$ GeV) in
order to be in agreement with the $pA$ measurement, and assumed the relation
\be
\sigma^{NUC}_{{J/\psi} N}(s) = \sigma^{NUC}_{{J/\psi} N}(s_0) 
\,\left(s/s_0\right)^\lambda
\label{abscross}
\ee
with $\lambda = 0.12$ in order to simulate nuclear effects as predicted in
\cite{KTH01}.

\section{Thermodynamics of the Produced Medium}
\label{sec_medium}
\subsection{Quasi-particle description of the QGP and the Hadronic EoS}
\label{sec_quasiparticles}

Before we address the question of how to describe the medium produced 
by a nuclear collision, we need to identify its relevant degrees of freedom.
On the theoretical side, information about the QCD medium comes from
lattice calculations at finite temperature. Simulations of the
pressure, entropy and energy density \cite{lat1} show a drastic
increase in the number of degrees of freedom at a critical temperature
of about 170 MeV (for two massless flavors \cite{lat2}) to about
80\% of the value of an ideal gas of quarks and gluons.  

A perturbative description of the QGP is, however futile, since
thermal perturbation theory is in general insufficient for all 
temperatures of interest. This is evident, for example, from
calculations of the free energy of the QGP \cite{ZHAI:1995} or the
photon self-energy in the thermal medium \cite{AURENCHE:1998}. 
Furthermore, close
to $T_c$ we expect intrinsically non-perturbative dynamics to enter:
the confinement/deconfinement transition and spontaneous 
chiral symmetry breaking are not accountable for in an expansion 
in the coupling constant. In view of these facts, we 
use a more phenomenological approach to QCD thermodynamics 
which nevertheless goes considerably beyond commonly used
ideal gas models. We have shown recently that it is possible to
describe the EoS of hot QCD to a very good
approximation by the EoS of a  gas of quasi-particles with thermally
generated masses, incorporating confinement effectively by a
temperature-dependent, reduced number of active degrees of
freedom. This reduction is caused by the formation of heavy hadrons or 
glueballs, with large masses compared to
$T_c$, and accounts for the fact that the QGP does not resemble at
all an ideal gas of quarks and gluons as the hadronization temperature
is approached. Here we give a short summary of the method and refer
the reader to \cite{SW01} for a more detailed discussion.

From asymptotic freedom, we expect that at extremely  high
temperatures the plasma consists of quasi-free quarks and gluons. Hard
thermal loop (HTL) perturbative calculations find, for thermal
momenta, spectral functions of the form $\delta(E^2 - k^2 - m^2(T))$
with $m(T) \sim g T$ \cite{LB96}. As long as the spectral functions of 
the thermal excitations at lower temperatures resemble qualitatively this
asymptotic form, a quasi-particle description is expected to be
applicable. QCD dynamics is then incorporated in the thermal
masses of the effective quarks and gluons, plus an extra function, 
required by thermodynamical consistency, which plays the role of the 
thermal vacuum energy.

The thermal excitations can then be described by a dispersion equation
\be
E^2_i(k,T) = \vec k^{\,2} + m^2_i(T)\,,
\label{dispersion}
\ee
where the subscript $i =g,q$ labels gluons
and quarks, respectively.
The thermal masses $m_i(T)$ are obtained from the self-energy of the
corresponding particle, evaluated at thermal momenta 
$k \sim T$. Eq.~(\ref{dispersion}) is valid as long as this self-energy
is only weakly momentum dependent in the relevant  kinematic region. 
In addition, for a quasi-particle to be a meaningful concept, we require its
thermal width to be small.  The quasi-particle mass, for 
instance of gluons, is then
\be
m_g(T) = T\ \sqrt{\frac{N_c}{6} + \frac{N_f}{12}} 
\ \tilde{g}(T, N_c,N_f) \label{m_gluon} 
\ee
with the effective coupling specified as
\be
\tilde{g}(T, N_c, N_f) = \frac{g_0}{\sqrt{11N_c - 2N_f}}
\left( [1 + \delta] - \frac{T_c}{T}
\right)^{\gamma}. \label{g_tilde}
\ee
$N_c$ and $N_f$ stand for the number of colors and flavors,
respectively. The functional dependence of $m_g(T)$ on $T$ is based on
the conjecture that the phase transition is second order or weakly 
first order which suggests an almost power-like behavior $m \sim (T -
T_c)^\gamma$ with some pseudo-critical exponent $\gamma > 0$. Setting
$g_0 = 9.4$, $\delta = 10^{-6}$ and $\gamma = 0.1$, the effective mass,
as given in eq.~(\ref{m_gluon}), approaches the HTL result at high 
temperatures. The
thermal quark mass is modeled similarly. Note that in contrast to
previous quasi-particle models extrapolated from HTL calculations
\cite{PESHIER:2000}, the thermal masses used here {\em drop} as $T_c$
is approached from above. Obviously for $T\gg T_c$ the 
perturbative limit of $m_g(T)$ and $m_q(T)$ is recovered.

The density of the QGP  takes the form
\be
n(T) = \sum_{q_i=1}^{N_f} \int\!\! \frac{d^3k}{(2\pi)^2}\ f_{q_i}(k,T)
+ \int\!\! \frac{d^3k}{(2\pi)^2} \ f_g(k,T)\,,
\label{p3_def}
\ee
where the distribution functions for quarks and gluons are given by
\be
f_{q_i,g}(k,T) = \nu_{q,g}\, C(T)
\left\{\! \rule{0pt}{12pt}\exp\left[\rule{0pt}{11pt} 
E_{q_i,g}(k,T)/T \right] \pm 1\right\}^{-1}\!\!\!\!\!\!,
\label{distqg}
\ee
being $\nu_q = 12$ and $\nu_g = 16$ the number of spin, flavor and color
polarizations of quarks ($N_f = 3$) and gluons, while
\be
C(T) = C_0 \left[ (1 + \delta_c) - \frac{T_c}{T} \right]^{\gamma_c}
\label{conffactor}
\ee
is the {\em confinement factor}. In order to reproduce lattice QCD
thermodynamics for two light quarks and one heavy quark, the parameters 
take the values $C_0 = 1.16$, $\delta_c = 0.02$ and $\gamma_c = 0.29$.
Pressure, energy density and entropy density of the QGP follow accordingly. 

The driving force of the transition at $T_c$, the confinement process, is
physics that we must include phenomenologically. Below $T_c$, the relevant
degrees of freedom are pions and heavier hadrons. Approaching $T_c$
from below, deconfinement sets in and quark and gluon quasi-particles 
are liberated. Conversely, when
approaching the phase transition from above, the number of thermally
active degrees of freedom is reduced due to the onset of
confinement. As $T$ comes closer to $T_c$, an increasing number of
quasi-particles gets trapped into hadrons (or glueballs).
All confinement does on a large scale is to effectively reduce the
number of thermally active degrees of freedom as the temperature is lowered.
This effect can be included in the quasi-particle picture by modifying the
distribution functions by a temperature-dependent factor $C(T)$ as given 
in eq.~(\ref{conffactor}). Its explicit form is obtained by calculating the
entropy density of the QGP with masses as in eq.~(\ref{m_gluon}).
Dividing the lattice entropy density by this result yields $C(T)$. 
Having fixed all parameters by extrapolating available lattice results to
realistic zero temperature quark masses, the EoS for two light and a strange
quark can be constructed for $T > T_c$. 

Below the critical temperature, we employ the EoS of a non-interacting
hadronic resonance gas, including all particles up to 1.6 GeV mass.
A pion chemical potential is later introduced as a mean to populate the 
pion phase-space occupation and reproduce the corresponding observed yield.
Although this prescription is oversimplified, any improvement has no bearing
on the final result of this work. A matching with the high temperature part
of the EoS is achieved by smoothly switching off the hadronic degrees of 
freedom.
\begin{figure}[t]
\begin{center}
\epsfig{file=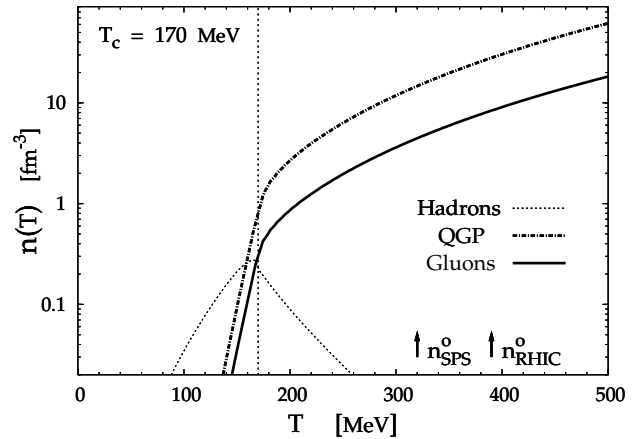,width=6cm, angle = -90}
\caption{Particle densities as a function of the temperature $T$ in
the  hadronic and partonic phases. Here
$n^0_{\rm SPS}$ denotes the initial quasi-particle gluon density at SPS and
$n^0_{\rm RHIC}$ the corresponding  value at RHIC. The vertical line
indicates the value of the critical temperature $T_c = $ 170 MeV.}
\label{figure:density}
\end{center}
\end{figure}

It is instructive to plot the total particle density given by
eq.~(\ref{p3_def}) as a function of temperature as shown in 
Fig.~\ref{figure:density}. The confinement
factor $C(T)$ accounts for the release of the constituents of
the hadrons as $T$ grows. Since it does not jump abruptly to one as
$T_c$ is reached, some hadronic clusters and hence non-zero hadron
densities persist slightly above the critical temperature, as well as 
quasi-particles persist slightly below $T_c$. 
We have indicated the values of particle densities at the beginning
and the end of the time evolution of the fireball, as will be discussed 
in the next section. Comparing the numbers involved, it is obvious that 
the number density in the QGP is always at least an order of magnitude
larger than the hadronic one (note the logarithmic scale). This has a 
crucial impact on $\Psi$ evolution in the produced medium.

\subsection{Time Evolution of the Fireball}
\label{sec_fireball}

Having characterized the matter which constitutes a strongly interacting 
system in the thermodynamic limit, we now address the intriguing issue
of how this system evolves in time after its production in a heavy-ion 
collision. Our main assumption is the following: the fireball reaches local 
thermal (though not necessarily chemical) equilibrium within a time scale of
the order of 1 fm/$c$. We then model the evolution dynamics 
by calculating the thermodynamic response of the hot and dense 
matter to the expansion of the total volume. Thermodynamical
parameters such as pressure $p$ and energy density $\epsilon$ in turn
feed back in the expansion dynamics. 

We average all quantities over the fireball volume, leaving us with
a spatially homogeneous system. The volume itself is taken to
be cylindrically symmetric around the beam (z-)axis. In order to
account for collective flow, we boost individual volume elements inside
the fireball volume with velocities depending on their position.
As flow velocities in longitudinal direction turn out to be close to the
speed of light, we have to include the effects of time dilatation.
On the other hand, we can neglect the additional time dilatation caused 
by transverse motion, for which the velocity is 
typically $v_\perp \ll v_z$. The thermodynamically relevant volume is
then given by the collection of volume elements corresponding to the
same proper time $\tau$. In order to characterize the volume expansion
within the given framework, we need first of all the expansion velocity
in longitudinal direction as it appears in the center of mass frame. 
Then we can compute the front position of the expanding cylinder as
\be
z(t) = v_z^0\, t + \int_{t_0}^{t} \!\! d t' \!\int_{t_0}^{t'} \!\!\!d t''
\ a_z(t'')\,,
\label{f1}
\ee
where $a_z(t) = c_z\, p(t)/\epsilon(t)$ is the longitudinal acceleration and
$c_z$ is a parameter.
The time $t$ starts running at $t_0$ such that $z_0 = v_z^0 \,t_0$ is
the initial longitudinal extension, $v_z^0$ being the initial longitudinal 
expansion velocity. The longitudinal position $z(t)$ and $t$ itself define 
a proper time curve $\tau = \sqrt{t^2 - z^2(t)}$. Solving for 
$\tilde t = t(\tau)$ one can construct $\tilde z(\tau) = z(\tilde t)$.
Then the position of the fireball front $z(t)$ in the center of mass frame 
is translated into the longitudinal extension $L(\tau$) of the cylinder
on the curve of constant proper time $\tau$. One obtains
\be
L(\tau) = 2\int_0^{\tilde z(\tau)}\!\!ds
\ \sqrt{1 + \frac{s}{\sqrt{s^2 + \tau^2}}}.
\label{f2}
\ee
At the same proper time we define the transverse flow velocity and 
construct the transverse extension of the cylinder as a disc of radius
\be
R(\tau) = R_0 + \int_{\tau_0}^{\tau} \!\! d \tau' \!\int_{\tau_0}^{\tau'} 
\!\!\!d \tau''\ a_\perp(\tau'')\,
\label{f3}
\ee
where $R_0$ is the initial overlap root mean square
radius of the two colliding
nuclei, $a_\perp(\tau) = c_\perp\, p(\tau)/\epsilon(\tau)$ is the transverse 
acceleration and $c_\perp$
is a parameter. With values of $L$ and $R$ obtained at a given
proper time, the 3-dimensional volume is parametrized as 
\be
V(\tau) = 2\pi R^2(\tau) \  L(\tau)\,,
\label{f4}
\ee
where the factor 2 arises from the relation $R_{Box} = \sqrt{2}\, R_{rms}$
between the radius of a sharp cylinder and its root mean square value.

The model is in principle fully constrained by fitting four parameters:
the initial longitudinal velocity $v_z^0$, the two constants $c_z$,
$c_\perp$ in $a_z$, $a_\perp$ and the freeze-out proper time
$\tau_f$. These are determined requiring
\be
R(\tau_f) \!=\! R_f,\ \, \frac{dR}{d\tau}(\tau_f) \!=\! v_\perp^f,\ \, 
\frac{d \tilde z}{d t}\left(\rule{0pt}{9pt}t(\tau_f)\right) 
\!=\! v_z^f,\ \, T(\tau_f) \!=\! T_f,
\ee
where $R_f$, $v_\perp^f$, $v_z^f$ and $T_f$ are extracted from
an appropriate set of experimental data. In practice this is achieved
only at SPS energy for central $Pb+Pb$ collisions, where the freeze-out 
analysis \cite{FREEZE-OUT} allows a complete determination. We then
assume entropy conservation during the expansion phase, fixing
the entropy per baryon $s_0$ from the number of produced
particles per unit rapidity. 
Calculating the number of participant baryons as 
\be
N_p(b) = \!\int\!\! d^2 r\ T_A(r)\! 
\left\{1-\exp\left[-\sigma^{in}_{pp}\,T_B(\tilde r)
\rule{0pt}{11pt} \right]\!\right\} \!+\! (A \leftrightarrow B),
\label{participants} 
\ee
where $\sigma^{in}_{pp} = 30$ mb at $\sqrt{s} = 17.3$ GeV (SPS) and 
$\sigma^{in}_{pp} = 42$ mb at $\sqrt{s} = 200$ GeV (RHIC), and scaling
the entropy with the calculated number of participants, we find 
$S_0 = s_0\,N_p$, the total entropy. The entropy density at a given proper time
is then determined by $s(\tau)=S_0/V(\tau)$. Using the EoS given by the 
quasi-particle approach, thereby providing constraints from lattice QCD, we 
find $T(s(\tau))$ and also $p(\tau)$ and $\epsilon(\tau)$. The set of
equations (\ref{f1}-\ref{f4}) is then solved iteratively, keeping the
entropy constant.
Finally we arrive at a thermodynamically self-consistent 
model for the fireball which is, by construction, able to reproduce the
hadronic momentum spectra at freeze-out.
This framework was shown to be a key ingredient for a successful description
of the low- and intermediate-mass dilepton yields in SPS collisions 
(see \cite{RSW02} where the model is also described in more detail).

While the fireball evolution is fully
constrained by data for central collisions at SPS, an extension towards 
different impact parameters and
higher beam energies is not on the same firm ground.
Unfortunately, no detailed freeze-out analysis has been done so far 
for different centralities, even at SPS. One has to rely on sensible 
assumptions in order to construct the fireball evolution.

Concerning the impact parameter dependence, from geometric considerations
one can compute the overlap area and convert it, for simplicity, to a disc,
while maintaining cylindrical symmetry but ignoring all effects of elliptic
flow. This approximation should still be valid for not too large impact 
parameters. Note that the model is not applicable for very peripheral events,
since the assumption of thermalization ceases to be valid.
We then make the following two assumptions: first,
the initial velocity $v_z^0$ of the fireball front is an important
quantity which must change for more peripheral collisions. We consider the
empirical fact that in $pp$ collisions leading particles
loose on average about one unit of rapidity. This is the limit we expect for
very peripheral collisions. In order to account for this effect,
we assume that the rapidity loss from incoming nuclei to the
bulk of the produced matter scales with the number of binary
collisions per participant ($\sim$ 2.7 for central $AB$ collisions, 
1 for $pp$) and interpolate linearly the value of $v_z^0$ between these 
limits. Second, we assume that the freeze-out temperature $T_f$ remains 
unchanged for different impact parameters. Since the total entropy scales 
with the number of participants $N_p$, all parameters are fixed and the 
fireball evolution can be generalized to non-central collisions.

The extension of the model to higher beam energies poses a more difficult
task. The main reason lies in the transition from a baryon-rich to a 
baryon-poor scenario. While such an analysis is certainly possible with 
the amount of data collected by RHIC so far, it is a difficult task
and has not yet been carried out. For the time being, we aim at a qualitative
description and refrain from giving detailed predictions. 
Assuming that the entropy per participant scales with the total
multiplicity, we can calculate the initial entropy $S_0$. 
In order to account for the decreasing baryon density
inside the fireball, we assume a rise in the freeze-out
temperature from 100 to 130 MeV when going from SPS to RHIC energy.
Similarly, we change the EoS in the hadronic part 
to incorporate the decreasing pion chemical potential with increasing energy
(for a more detailed discussion, see \cite{RSW02}). The thermalization time 
decreases from 1 fm/c at SPS to 0.6 fm/c at RHIC.
From $dN/dy$ particle distributions we extract the rapidity extent of
hadron spectra to be of 5.5 units, allowing $v_z^f$ to be extracted.
Moreover, experimental results indicate that transverse flow and
the geometrical freeze-out radius are virtually unchanged from
SPS to RHIC energies \cite{HBT}, so we keep these two parameters
unchanged from the SPS case. Finally we adjust $v_z^0$ consistently with the 
assumed freeze-out temperature. 
\begin{table}[b]
\begin{center}
\begin{tabular}{|c|c|c|}
\hline
     & SPS $Pb\!+\!Pb$ & RHIC $Au\!+\!Au$\\
     & $\sqrt{s}$ = 17.3 GeV & $\sqrt{s}$ = 200 GeV \\
\hline
\hline
   \hspace{1mm} $\tau_0$ \hspace{1mm} & \hspace{10mm} 1.0 [fm/$c$] 
	\hspace{1mm}  & \hspace{11mm} 0.6 [fm/$c$] \hspace{1mm} \\
\hline
   \hspace{1mm} $\tau_c$ \hspace{1mm}  & 8.0 & 8.7 \\
\hline
   \hspace{1mm} $\tau_f$ \hspace{1mm}  & 18 & 17 \\
\hline
   \hspace{1mm} $T_0$ \hspace{1mm} & \hspace{9mm} 320 [MeV] 
     & \hspace{8.5mm} 390 [MeV] \\
\hline
   \hspace{1mm} $T_f$ \hspace{1mm}  & 100 & 130 \\
\hline
   \hspace{1mm} $R_0$    \hspace{1mm}  & \hspace{5mm} 4.8 [fm] 
     & \hspace{5mm} 4.8 [fm] \\
\hline
   \hspace{1mm} $R_c$    \hspace{1mm}  & 5.2 & 6.0 \\
\hline
   \hspace{1mm} $R_f$    \hspace{1mm}  & 9.0 & 9.0 \\
\hline
\end{tabular}
\end{center}
\caption{\label{T-Fireball}Key properties of the fireball evolution 
deduced for central collisions at SPS and RHIC. The numbers given for the 
radii are root mean square values.}
\end{table}

\begin{figure}[t]
\begin{center}
\epsfig{file=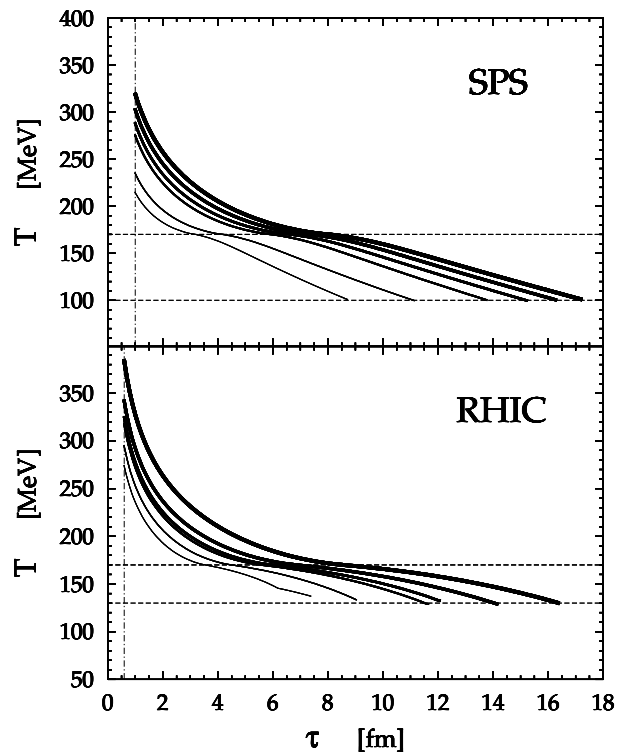,width=8.5cm,angle=0}
\caption{Temperature $T$ as function of proper time $\tau$ for various 
collisions at SPS and RHIC energies. Curves correspond to impact parameters
$b =$ 0-1 fm (thicker curves), 2-3, 4-5, 6-7, 8-9 and 10-11 fm (thinner
curves).}
\label{figure:fireball}
\end{center}
\end{figure}
The model parameters are eventually determined with the self-consistent
procedure described above. Relevant 
values of parameters for central collisions at SPS and RHIC are reported 
in Tab.~\ref{T-Fireball}. One notices the significant QGP lifetime of
about 8 fm, almost half of the total lifetime of the fireball. Temperature 
$T$ profiles as function of proper time $\tau$ for collisions at different
impact parameters at SPS and RHIC energies are plotted in 
Fig.~\ref{figure:fireball}. Note the delayed cooling occurring at the 
transition temperature $T_c =  170$ MeV. This is due to the softening of
the EoS, influencing the volume, and therefore temperature, expansion 
pattern via the acceleration $a(\tau) \propto p(\tau)/
\epsilon(\tau)$ in eqs.~(\ref{f1}) and (\ref{f3}).
The presence of a longitudinal acceleration term also
leads to interesting consequences. The smaller initial rapidity distribution
of matter implies a larger initial energy density and hence a higher
temperature of the fireball as compared to the Bjorken estimate for
these quantities.

Particle densities at the corresponding initial temperatures for central 
collisions at SPS and RHIC are indicated in Fig.~\ref{figure:density} in
the previous section.

\section{Charmonium Evolution in the Expanding Matter}

\subsection{Interaction of {\boldmath $\Psi$} with the produced medium}

After having discussed the initial conditions for open and hidden charm
production in $AB$ collisions and the physics of the expanding medium,
we now come to the central issue of the paper and examine the
interaction between charmonium and the produced medium. Because of the 
relatively small value of the ${J/\psi}$ formation time
\be
\tau_f^\Psi \sim (m_{J/\psi} - m_{\psi'})^{-1} \sim 0.3\ {\rm fm}\,,
\ee
in the following we will assume that ${J/\psi}$s are fully formed hadrons
by the time the thermalized medium is produced, and they will subsequently
interact with its degrees of freedom. 
Since the medium itself consists of a QGP for a significantly long time, 
we begin discussing how ${J/\psi}$ interacts with quark and gluon 
quasi-particles.
As mentioned in the Introduction, we neglect the possibility of static Debye 
screening, thereby focusing our analysis on the collisions of $\Psi$ with
the QGP degrees of freedom. We argue that a proper treatment of screening
and collisions (together, in order to avoid double counting) is a complex 
dynamical problem involving an unknown time-scale for the modification of 
the $c \bar c$ binding potential. We therefore refrain from speculating on 
this issue, which deserves a separate study, and simply assume that 
$\Psi$s produced by the initial hard processes will propagate and collide
with the QGP constituents.
  
\begin{figure}[t]
\begin{center}
\epsfig{file=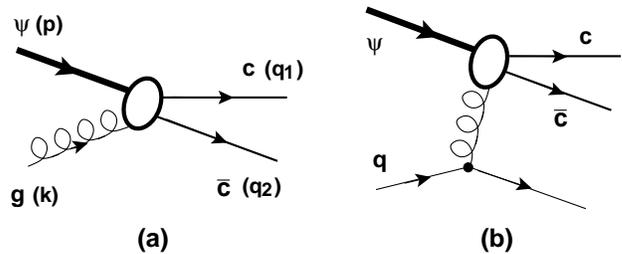,width=8.5cm}
\caption{Diagrams contributing to lowest order to $\Psi$
dissociation. The  process involving a gluon quasi-particle (a)
exhausts the amplitude since it already takes into account the one
involving a quark quasi-particle (b).}
\label{figure:diagrams}
\end{center}
\end{figure}

It is clear that collisions of $\Psi$ with either quarks or gluons will lead
to dissociation of the bound state.
One expects that the two processes illustrated in Fig.~(\ref{figure:diagrams})
contribute to $\Psi$ dissociation in leading order. 
On the other hand, a quark can interact only
via  gluon exchange. Within the quasi-particle model, the
process labelled (b) in the figure is effectively already included in
the definition of the temperature dependent gluon mass. Computing both
contributions would lead to an erroneous double counting. In other
words, $\Psi$s only see gluonic quasi-particles in the plasma. 
As an exception, processes such as $\Psi c \rightarrow g c$ and $\Psi
\bar c \rightarrow g \bar c$ could be important. For example, the first 
one can proceed by dissociating $\Psi \rightarrow c \bar c$ and 
annihilating the $\bar c$ with an incoming $c$ to form a gluon.
Nevertheless such processes can be neglected when compared to gluon
dissociation due to the much larger density of the gluons themselves.

We concentrate now on the process (a) in
Fig.~(\ref{figure:diagrams}) for the reaction $\Psi(p)+g(k) \rightarrow 
c(q_1) +\bar c(q_2)$, and label momenta as indicated in parentheses.
We then come to the problem of computing 
cross sections involving a relativistic bound state. In the present
case one can argue that the $c \bar c$ system is, to first
approximation, non-relativistic, greatly simplifying the treatment.
Moreover, as was done originally by Bhanot and Peskin \cite{P79,BP79}
one can argue that the lowest-lying quarkonium levels can be
approximately described using the Coulomb part of the potential. 
Then, with operator product expansion methods or more recent non-relativistic
factorization techniques \cite{OKL01}, it is possible  to obtain an
analytic expression for the cross section. For the case of $J/\psi$
the result is
\be 
\sigma_{J/\psi\,g}(\omega) = A_0
\,\frac{\left(\omega/\epsilon_{J/\psi}
-1\right)^{3/2}}{\left(\omega/\epsilon_{J/\psi}\right)^5}\,,
\label{bpcross}
\ee 
where $A_0 = (2^{11}\pi/27)\,(m_c^3\,\epsilon^{\,0}_{J/\psi})^{-1/2}$.
The cross section is a
function of the gluon energy $\omega$ in the $J/\psi$ rest frame
and involves the threshold energy $\epsilon_{J/\psi}$,
related to the binding energy $\epsilon^{\,0}_{J/\psi}$ by the condition
$(p+k)^2 \geq 4m_c^2$, implying that
$\epsilon_{J/\psi} = 
\epsilon^{\,0}_{J/\psi}+(\epsilon^{\,0}_{J/\psi})^2/(2 m_{J/\psi})$.  
The binding energy is taken to be $\epsilon^{\,0}_{J/\psi} = 800$
MeV and the charm quark mass $m_c = 1.95$ GeV as in
\cite{BP79}, to fit the masses of first two charmonium levels 
($m_{J/\psi} = 3.1$ GeV and $m_{\psi'} = 3.7$ GeV).
In an analogous fashion, it is possible to compute the cross section for 
$\psi'$ dissociation by gluons. Given the binding energy 
$\epsilon^{\,0}_{\psi'} = \epsilon^{\,0}_{J/\psi}/4$ (for a Coulombic system),
and the cross section is
\be 
\sigma_{\psi' g}(\omega) = 16\, A_0
\,\frac{\left(\omega/\epsilon_{\psi'}
-1\right)^{3/2} \left(\omega/\epsilon_{\psi'}
-3\right)^2}{\left(\omega/\epsilon_{\psi'}\right)^7}\,.
\label{bpcross2}
\ee 
Recently also the dissociation cross section for $\chi_c$ states
was computed \cite{WY02}. With a mass $m_{\chi_c} = 3.5$ GeV and a binding 
energy $\epsilon_{\chi_c} = 400$ MeV one obtains
\be 
\sigma_{\chi_c g}(\omega) = 4 A_0
\frac{\!\!\left(\omega/\epsilon_{\chi_c}
-1\right)^{1/2} \!
\left[\rule{0pt}{10pt}
9(\omega/\epsilon_{\chi_c})^2\!\! -\! 20(\omega/\epsilon_{\chi_c})\! +\! 12  
\right]}
{\left(\omega/\epsilon_{\chi_c}\right)^7}\!.
\label{bpcross3}
\ee 

\begin{figure}[t]
\begin{center}
\epsfig{file=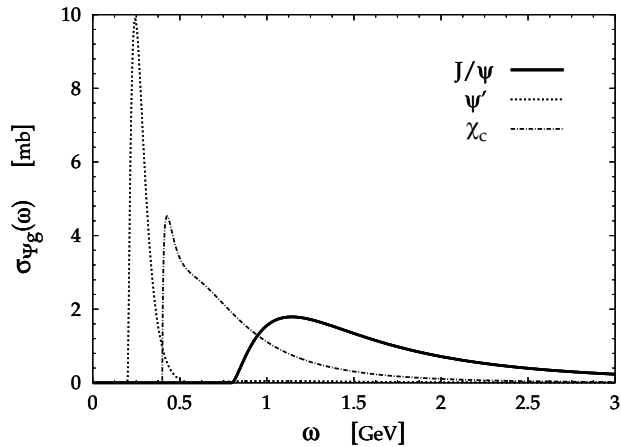,width=6cm,angle=-90}
\caption{Dissociation cross sections for the processes
$\Psi\,g \rightarrow c \bar c$, where $\Psi = J/\psi$,
$\psi'$ and $\chi_c$, as function
of the gluon energy in the rest frame of $\Psi$.}
\label{figure:bpdiss}
\end{center}
\end{figure}
The cross sections calculated with eqs.~(\ref{bpcross}), (\ref{bpcross2}) and
(\ref{bpcross3}) are not necessarily reliable since the Coulomb potential
approximation is used without justification. This problem is serious
especially for the $\psi'$ and $\chi_c$ states, which have much too large
dissociation cross sections. Although the inclusion of
higher states is in principle necessary in order to have a reliable
description of the dissociation process, since $\psi'$ and $\chi_c$
constitute about 40$\,$\% of the final $J/\psi$ yield, it is not possible
at this stage to pursue further this approach. A way out could be to compute
these cross section using a realistic potential for the $c \bar c$ system,
but then questions arise for a rigorous separation of perturbative and
non-perturbative contributions and such a study falls  beyond the scope of
the present treatment. Within the adopted framework we choose to re-scale
the $\psi'$ and $\chi_c$ cross sections by a multiplicative factor $\kappa$
such that the obtained $J/\psi$ suppression pattern at SPS energies is
in agreement with experiment. The required value is $\kappa = 0.2$ and the
resulting cross sections are plotted in Fig.~(\ref{figure:bpdiss}).
One notices the great difference in magnitude between the $J/\psi$ case 
as compared to $\psi'$ and $\chi_c$, even after re-scaling. We want to stress
the fact that the value $\kappa = 0.2$ is not necessarily large. In fact
cross sections are proportional to the square of a matrix element. In the
present case, since a gluon is a vector particle, the matrix element consists
in a derivative of the $\Psi$ wave function in momentum space. A realistic
confining potential would squeeze the spatial wave function more than the
Coulomb potential does. In momentum space the wave function would then be 
more extended and flatter, resulting in a much lower cross section.
Within the adopted framework we denote with $\sigma_{\Psi g \rightarrow c \bar
c}(s)$ the $\Psi g$ dissociation cross section, the dependent variable
being the center of mass energy $s$, related to the gluon energy in the $\Psi$
rest frame as $\omega = k_\mu p^\mu/m_\Psi = (s - p^2 -k^2)/(2\,m_\Psi)$.

\begin{figure}[t]
\begin{center}
\epsfig{file=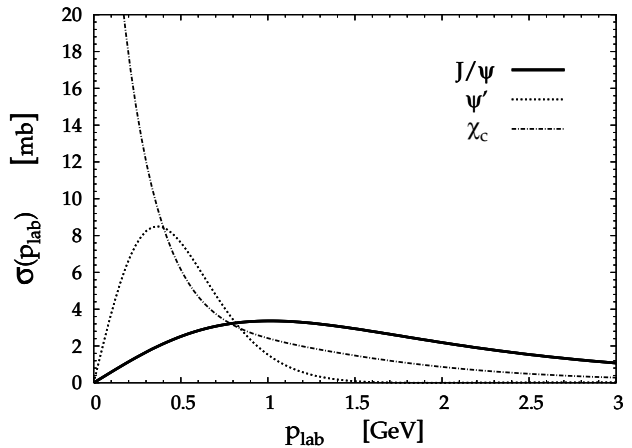,width=6cm,angle=-90}
\caption{Formation cross sections for the process
$c \bar c \rightarrow \Psi g$, where $\Psi = J/\psi$,
$\psi'$ and $\chi_c$, according to 
eq.~(\ref{detbal}), as function of the laboratory momentum in the rest
frame of the target $\bar c$. }
\label{figure:dissform}
\end{center}
\end{figure}

We now consider the possibility of $c \bar c$ coalescence in the QGP,
a manner for production of $\Psi$ which has been recently considered
by several authors. Processes such as $g c \rightarrow \Psi c$ and $g
\bar c \rightarrow \Psi \bar c$ could in principle also contribute,
due to the large number of gluons available. On the other hand they
require the creation of a $c \bar c$ pair from the vacuum and are
therefore suppressed. In our formulation we then take into account
only the fusion process. We do this by means of a cross section,
applying detailed balance to the reaction $\Psi g
\leftrightarrow c \bar c$, and use the cross section calculated above
for $\Psi$ dissociation by gluons. Choosing the zero
momentum frame, flux factors are identical for the direct and reverse
processes and simple kinematics guarantees that the relation 
\be
\sigma_{c \bar c \rightarrow \Psi g}(s) = \sigma_{\Psi g
\rightarrow c \bar c}(s) \ \frac{4}{3}\, \frac{(s - m_\Psi^2)^2}{s\,(s
- 4m_c^2)}
\label{detbal}
\ee 
holds. The factor $4/3$ arises from counting
the number of degrees  of freedom (spin and color factors) in the two
different channels. For the $\Psi g$ system they are $(1 \times 3)
\times (8 \times 2) = 48$, while for the $c \bar c$ system $(3 \times 2)
\times (3 \times 2) = 36$. Since $s = 2m_c(E_c+m_c)$ for the $c \bar c$
system, one can express  eq.~(\ref{detbal}) as function of the $c$
quark momentum $p_{lab}$ in the rest frame of the $\bar c$. The result for
$\sigma_{c \bar c \rightarrow \Psi g}$ is given in 
Fig.~(\ref{figure:dissform}), where the formation cross sections for
$J/\psi$, $\psi'$ and $\chi_c$ are shown.

Coming now to the case of hadronic dissociation, we recall that many 
approaches have been developed in the literature to compute mesonic
dissociation cross sections of $\Psi$. The first calculations were
presented in \cite{MBQ95}. Subsequently, more sophisticated treatments
were developed (See \cite{WSB} as an example). The resulting thermally 
averaged cross section amounts to a few mb for the process $\Psi \rho
\rightarrow D \bar D$, but computations considered $\rho$ to be a particle
with infinite life-time, neglecting its broad spectral function. Accounting
for this should considerably reduce the value of the calculated cross section.
Significantly smaller values are obtained for
$\Psi \pi \rightarrow D \bar D$. Despite the efforts applied to perform
the mentioned calculations, a realistic 
time evolution of the fireball was never employed to obtain results comparable
with experimental data. We stress that a proper and independent description  
of the produced medium is crucial if one wants to make use of charmonia
as probes of dense matter. In our approach the medium evolution is 
constrained by the freeze-out analysis, leaving no room for adjustments.
According to the calculation of the particle density as function of 
temperature, plotted in Fig.~(\ref{figure:density}) in the previous
section, we see that particle densities in the hadronic phase are almost
two orders of magnitude lower than in the QGP phase. Note the logarithmic
vertical scale. Unless one employs exceptionally large cross sections,
it seems unlikely that hadronic dissociation can be at all relevant.
A way out was proposed in \cite{BBK01}, where it was argued that the 
QCD analogue of the Mott transition in an electron plasma may produce
an onset behavior in thermally averaged cross section when the
Mott temperature $T_{\rm Mott} \sim T_c$ is reached. However, this 
framework has not yet been constrained by lattice EoS results.
Given the inherent uncertainties and the above remarks on the
smallness of particle densities in the hadronic phase, we will not
consider the possibility of hadronic dissociation of $\Psi$.

\subsection{Kinetic Description of Dissociation and Formation}

After having discussed and characterized the medium produced in a heavy-ion
collision and specified how charmonium interacts by collisions with its
constituents, we are ready to address the problem of how produced charmonia
evolve and propagate through the medium until their final experimental
observation.
The natural framework in which to study the time evolution of $\Psi$ is that
of kinetic theory. We use a semi-classical treatment, setting up a
relativistic kinetic equation for the $\Psi$ phase-space distribution.
The general problem of kinetic evolution of bound states in a strongly
interacting medium was studied recently \cite{P03}. Here we make use of 
those results, adapting the treatment to our needs. We then assume that
the kinetic equation  describing the time evolution of
the phase-space density $\fp$ of $\Psi$ is
\be
p^\mu \partial_\mu\, \fp(p,x) = C^\Psi_F(p,x) - C^\Psi_D(p,x)\,\fp(p,x)\,,
\label{equation}
\ee
consisting of the drift term $p^\mu \partial_\mu\,\fp$ on the l.h.s. and of
a collision term on the r.h.s. The collision term is made of a 
dissociation (loss) part
\bea
C^\Psi_D(p,x) \!\!&=&\!\! \frac{1}{2} \int\! d\Phi_3(k,q_1,q_2)
\ (2\pi)^4 \delta^4(p\!+\!k\!-\!q_1\!-\!q_2) \nonumber \\
& &\hspace{1cm} \times\ {\overline W}_{\Psi g \rightarrow c \bar c}
\ f_g(k,x) 
\label{colld}
\eea
describing $\Psi$ dissociation by quasi-particle gluons, studied in 
\cite{BO89,XKSW96} to address the problem of charmonium suppression,
and a formation (gain) part
\bea
C^\Psi_F(p,x) \!\!&=&\!\! \frac{1}{2} \int\! d\Phi_3(k,q_1,q_2) 
\ (2\pi)^4 \delta^4(p\!+\!k\!-\!q_1\!-\!q_2) \nonumber \\
& &\hspace{1cm} \times\ {\overline W}_{c \bar c \rightarrow \Psi g}
\ f_{c}(q_1,x)\,f_{\bar c}(q_2,x) 
\label{collf}
\eea
describing $\Psi$ formation by $c \bar c$ fusion, introduced in a simplified
manner in \cite{TSR01}. We stress here that 
eq.~(\ref{equation}) stands actually for 3 equations, for $J/\psi$, $\psi'$
and $\chi_c$.

In the equations above $f_{c}, f_{\bar c}$ and $f_g$ are the phase-space 
distributions of the degrees of freedom participating in the collisions.
Note that effects of Bose enhancement for gluons and Pauli blocking for $c$
and $\bar c$ quarks are negligible. The space-time 4-vector is
denoted by $x$, while kinematics is the same as in the previous section, 
that is $p$ is the 4-momentum of $\Psi$, $k$ is that of the gluon, while
$q_1$ and $q_2$ are those of the $c$ and $\bar c$ quarks. The Lorentz
invariant 3-body the phase-space integration measure is
\be
d\Phi_3(k,q_1,q_2)  = 
\frac{d^3 k}{(2\pi)^3 2E_k}\frac{d^3 q_1}{(2\pi)^3 2E_1}
\frac{d^3 q_2}{(2\pi)^3 2E_2}\,.
\label{ph3}
\ee
All particles are assumed to be on their mass shells. 

The transition probabilities ${\overline W}$ are averaged over the initial
color and spin polarizations and summed over the final ones. Without 
the bar $W$ indicates a transition probability already summed over
both initial and final polarizations. Therefore 
${\overline W}_{c \bar c \rightarrow \Psi g} = N_{c \bar c}^{-1}\,
W_{c \bar c \rightarrow \Psi g}$ and 
${\overline W}_{\Psi g \rightarrow c \bar c} = N_{\Psi g}^{-1}\,
W_{\Psi g \rightarrow c \bar c}$, with $N_{c \bar c} = 36$ and 
$N_{\Psi g} = 48$. The transition probabilities satisfy detailed balance as
\be
W_{c \bar c \rightarrow \Psi g} = W_{\Psi g \rightarrow c \bar c}\,.
\label{detbalw}
\ee
They are related to the cross sections discussed in the previous section,
which can be expressed as
\be
\sigma^\Psi_D(s) \!=\! \frac{1}{\!4F_{\Psi g}}\!\!\int\!\! d\Phi_2(q_1,q_2)
 (2\pi)^4\! \delta^4(p\!+\!k\!-\!q_1\!-\!q_2)
\,{\overline W}_{\Psi g \rightarrow c \bar c}
\label{defdiss}
\ee
for dissociation and
\be
\sigma^\Psi_F(s) \!=\! \frac{1}{4F_{c \bar c}}\!\int\! d\Phi_2(p,k)
\, (2\pi)^4 \delta^4(p\!+\!k\!-\!q_1\!-\!q_2)
\ {\overline W}_{c \bar c \rightarrow \Psi g}
\label{defform}
\ee
for formation. Here the 2-body phase-space integration measure is
\be
d\Phi_2(p_a,p_b)  = \frac{d^3 p_a}{(2\pi)^3 2E_a}
\frac{d^3 p_b}{(2\pi)^3 2E_b}\,,
\ee
while the flux factors are $F_{cd} = \sqrt{(p_c\cdot p_d)^2 - m_c^2 m_d^2 }$.
In the non-relativistic approximation used in \cite{OKL01}, the transition
probabilities depend only on the center of mass energy $s$. This allows
to relate them to the cross sections as
\be
\sigma_{ab \rightarrow cd}(s) = \frac{1}{16 \pi\, s}
\,\frac{F_{cd}(s)}{F_{ab}(s)}\ {\overline W}_{ab \rightarrow cd}\,.
\label{crossw}
\ee
It is then trivial to see that eqs.~(\ref{detbalw}), (\ref{defdiss})
and (\ref{defform}) are consistent with eq.~(\ref{detbal}).

Concerning the definitions of the phase-space distribution of gluons, we
adopt the expression given in eq.~(\ref{distqg}) of section 
\ref{sec_quasiparticles}. The time evolution of the gluon distribution is
all contained in the proper time dependence of the temperature. For $c$ 
quarks, as they are produced in the hard initial collision, we take their
spectrum to be given by eq.~(\ref{charmAB}). We assume that they do not
interact significantly with the medium, approximately moving on straight
lines according to the free streaming equation 
\be
q^\mu \partial_\mu\, f_{c, \bar c}(q,x) = 0\,.
\ee
The initial spatial distribution is assumed to be very narrow. Introducing
for convenience the space-time rapidity variable $\eta = 1/2\, 
\log[\,(t+z)/(t-z)\,]$, we define the phase-space density of charm quarks as
\be
f_{c}(q,x) = \frac{(2 \pi)^3}{\tau\, m_\perp^c}
\,\frac{dN^c_{AB}}{dy_c d^2\vec q_\perp}
\ \delta(y_c\!-\!\eta)\, \rho_\perp\!\!\left(\rule{0pt}{9pt}r\!_\perp(\tau)
\right) \theta\!\left(\rule{0pt}{9pt}T(\tau) \!-\! T_c\right) 
\label{cdist}
\ee
and analogously for $\bar c$ quarks. The step function $\theta$ is introduced
to account for the fact that below $T_c$ $c$ and $\bar c$ quarks have 
hadronized into $D$ mesons and are no more available to coalesce into $\Psi$.
For simplicity in later numerical computations, the transverse position 
density is taken as a box of the same radius of the fireball as
\be
\rho_\perp\!\!\left(\rule{0pt}{9pt}r\!_\perp(\tau) \right) =
\frac{1}{\pi R^2(\tau)}
\,\theta\left(\rule{0pt}{9pt}R(\tau) - r_\perp(\tau)\right),
\label{transbox}
\ee
where
$\vec r_\perp(\tau) = \vec r_\perp - (\vec q_\perp/m_\perp^c)\,\tau$. 
The $\delta$-function, arising from the assumption that $c$ quarks are
produced in a very narrow longitudinal region, strongly correlates
$y_c$ and $\eta$. This is a strong constraint on $c \bar c$ coalescence
and we will discuss its implications later on.

All the dynamics is now well defined since we have specified the 
distributions of the medium constituents and those of the bound state 
constituents, together with the transition probabilities 
$W_{\Psi g \leftrightarrow c \bar c}$. The kinetic equation can 
be solved exactly in closed form \cite{P03}. However, the medium evolution
has been constructed by averaging over the whole interaction volume. It is
therefore not possible to exploit the full phase-space information contained
in eq.~(\ref{equation}). We therefore simplify the kinetic equation as
follows. 

We recall that the $\Psi$ momentum spectrum at a given constant proper time 
can be computed from the phase-space distribution $\fp$ by means of the 
Cooper-Frye formula \cite{CF74}. Integrating out also the transverse
momenta, we obtain the $\Psi$ rapidity distribution
\be
\frac{dN_\Psi}{dy}(\tau) = \frac{1}{\!(2 \pi)^3}
\!\!\int\!\!d^2 p_\perp \!\!\int\!\! d^2 r_\perp d\eta
\ \tau\, m_\perp^\Psi \cosh(y\!-\!\eta)\ \fp(p,x).
\label{psirapdist}
\ee
If we now re-express the operator $p^\mu \partial_\mu$ appearing in
eq.~(\ref{equation}) by means of the variables $y$ and $\eta$, we
obtain
\be
p^\mu \partial_\mu \!= m_\perp\!\! \left[ \cosh(y\!-\!\eta)
\frac{\partial}{\partial \tau} \!+\!
\frac{1}{\tau} \sinh(y\!-\!\eta) \frac{\partial}{\partial \eta} \right]
+ \vec p_\perp \!\cdot\! \frac{\partial}{\partial \vec r_\perp}.
\ee
It is now simple to see that differentiating with respect to $\tau$ the 
$\Psi$ spectrum given with eq.~(\ref{psirapdist}) is equivalent to integrating
the l.h.s. of eq.~(\ref{equation}) in all phase-space variables except $y$.
More precisely, we obtain
\be
\frac{\partial}{\partial \tau}\frac{dN_\Psi}{dy}(\tau) =
\frac{1}{(2 \pi)^3}\!\int\! d^2 p_\perp  
\!\int\!d^2 r_\perp d\eta\ \tau\,
 \ p^\mu \partial_\mu\, \fp(p,x).
\label{simple}
\ee

Integrating the r.h.s. of eq.~(\ref{equation}) is a bit more complicated.
We first discuss the dissociation term and make use of its definition
in eq.~(\ref{colld}) and of the definition of dissociation cross section
given in eq.~(\ref{defdiss}) to write
\be
C^\Psi_D(p,x) = \int \!\! \frac{d^3 k}{(2\pi)^3 E_k}\, F_{\Psi g}(s)
\,\sigma^\Psi_D(s)\, f_g(k,x)\, \fp(p,x).
\ee
If we now integrate $C^\Psi_D$ over all phase-space variables except $y$ we
obtain
\bea
\int\! d^2 p_\perp  d^2 r_\perp d\eta\ \tau
\ C^\Psi_D(p,x)\, \fp(p,x) \ \simeq \hspace{2.5cm}
&& \nonumber \\
&&\hspace{-8.5cm}
\simeq \int \!\! \frac{d^3 k}{(2\pi)^3} 
\,V_{\Psi g}\,\sigma^\Psi_D(s)\,f_g\!\left(k,T(\tau)\rule{0pt}{9pt}\right)
\!\int\! d^2 r_\perp d\eta\  \tau  E_\Psi \, \fp(p,x) \nonumber \\
&&\hspace{-8.5cm}
\simeq\ \lambda^\Psi_D(\tau)\  \frac{dN_\Psi}{dy}(\tau), \label{step2}
\eea
where the dissociation rate is defined as
\be
\lambda^\Psi_D(\tau) = \left. \int \!\! \frac{d^3 k}{(2\pi)^3} 
\,V_{\Psi g}\ \sigma^\Psi_D(s)\ f_g\!\left(k,T(\tau)\rule{0pt}{9pt}\right)
\  \right|_{\begin{array}{l} _{y\ =\ 0}\ \ \ ^{\displaystyle .} \\ 
^{p_\perp\ =\ \langle p_\perp \rangle} \end{array}} \hspace{-5mm}
\label{rated}
\ee
\vspace{-4mm}

\begin{figure}[t]
\begin{center}
\epsfig{file=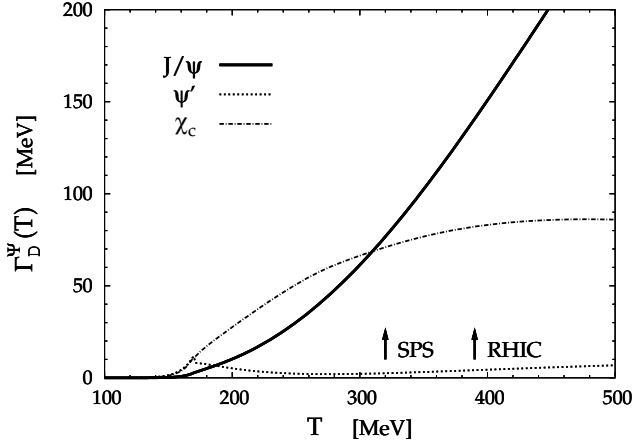,width=6cm,angle=-90}
\caption{Dissociation widths of $J/\psi$, $\psi'$ and $\chi_c$ as
function of the temperature. Arrows indicate initial temperatures for SPS and
RHIC central collisions.}
\label{figure:lambda1}
\end{center}
\end{figure}

\noindent To obtain our result we have made certain approximations.
First of all we have suppressed the argument $x$ in $f_g$ since it is 
redundant, the medium being uniform, and replaced it with the time
dependent temperature. The first step to arrive at eq.~(\ref{step2}) 
is obtained  approximating the $p_\perp$ dependence of $V_{\Psi g} = 
F_{\Psi g}(s)/E_\Psi E_k$ and $\sigma_D$ with its mean value. We
neglect for simplicity the dependence of the mean value on
centrality and choose $\langle p_\perp \rangle = 2.0$ GeV, whose
magnitude is in the range of experimental measurements, for all
calculations. Our results are not very sensitive to the precise
value of $\langle p_\perp \rangle$, especially after folding them with the
fireball time evolution. In the second step we assumed that
$\fp$ is peaked around values of $y$ close to $\eta$, as for the $c$ quark
distribution in eq.~(\ref{cdist}), and set $y=0$ which is the relevant
value for later comparing our final results with experiment. The
dissociation rate, after converting it to a dissociation width by
changing dimensions from fm$^{-1}$ to MeV, is plotted in 
Fig.~\ref{figure:lambda1}. There the temperature dependence for $J/\psi$,
$\psi'$ and $\chi_c$ is shown. One notices that the $J/\psi$ width is 
smaller than the one of $\chi_c$ for temperatures below $\sim 300$ MeV. 
$J/\psi$ becomes broader at higher temperatures, while the $\chi_c$
width stays virtually constant. Notice that the effect on $\psi'$ is
negligible. The different behaviors are related to the different forms and
thresholds of the dissociation cross sections, as illustrated in
Fig.~\ref{figure:bpdiss}. 

\begin{figure}[t]
\begin{center}
\epsfig{file=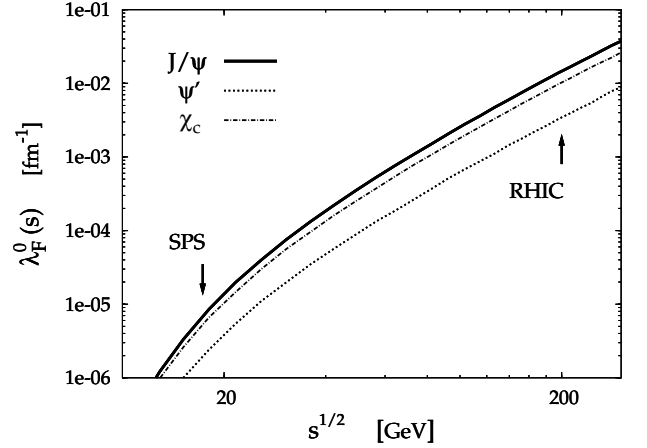,width=6cm,angle=-90}
\caption{Formation rates of $J/\psi$, $\psi'$ and $\chi_c$ computed at 
  the initial evolution time as function of the
  collision energy. The values increase by more than three orders of
  magnitude from SPS to RHIC central collisions.}
\label{figure:lambda2}
\end{center}
\end{figure}

Considering now the formation term on the r.h.s. of eq.~(\ref{equation}),
we integrate $C^\Psi_F$ in the same way as for $C^\Psi_D$ over the 
phase-space variables and define the formation rate
\be
\lambda^\Psi_F(\tau) = \left. \frac{1}{(2 \pi)^3} 
\int\! d^2 p_\perp \!\int\! d^2 r_\perp d\eta
\ \tau \ C^\Psi_F(p,x) 
\ \right|_{\, y \, =\,  0} \hspace{-2mm} .
 \hspace{-3mm}
\label{step3}
\ee
Its detailed evaluation proceeds without approximations but goes through
several steps. We leave all details in the Appendix and here report the
final result which is
\be
\lambda^\Psi_F(\tau) = \frac{1}{\tau}\!\int\! d^2 p_\perp
d^2 q^{\ 1}_\perp d^2 q^{\ 2}_\perp\, K\, \sigma^\Psi_F(s)
\, S_c(\bar y_c,q^1_\perp)S_c(\bar y_c,q^2_\perp)
\label{ratef}
\ee
where $K$ is a dimensionless quantity given in eq.~(\ref{kappa}), while 
$S_c(y,q_\perp)$ is the $c$ quark spectrum as given in eq.~(\ref{charmAB}).
The resulting formation rate $\lambda_F^0 = \lambda^\Psi_F(\tau_0)$ is
plotted for $J/\psi$, $\psi'$ and $\chi_c$
in Fig.~\ref{figure:lambda2} at the initial evolution time, as function
of the collision energy. The values increase by more than three orders of
magnitude from SPS to RHIC central collisions, thereby suggesting that
the mechanism of $\Psi$ formation by $c \bar c$ coalescence might be
important at high energies.

With all the simplifications mentioned before, we have managed to reduce 
eq.~(\ref{equation}) to a simple first order differential equation for the
rapidity distribution of $\Psi$ as function of proper time. In fact,
putting together eqs.~(\ref{simple}), (\ref{step2}) and (\ref{step3}) we
obtain that
\be
\frac{\partial}{\partial \tau}\frac{dN_\Psi}{dy}(\tau) =
\lambda^\Psi_F(y,\tau) - \lambda^\Psi_D(y,\tau)\, \frac{dN_\Psi}{dy}(\tau)\,,
\ee
whose solution, valid at $y = 0$, is obtained with few elementary steps 
and provides the final $\Psi$ rapidity distribution at $y=0$ as
\bea
\frac{dN^f_\Psi}{dy} \!&=& \!
\left\{\rule{0pt}{16pt} \frac{dN_\Psi^0}{dy}
\,\exp\left[ -\! \int_{\tau_0}^{\tau_f}\!\!\!d\tau'\,
\lambda^\Psi_D(y,\tau')\right] \right. \label{solution} \\
&& \left. \hspace{1mm} + \int_{\tau_0}^{\tau_f}\!\!\!d\tau'\, 
\lambda^\Psi_F(y,\tau')\, \exp\left[ -\! \int_{\tau'}^{\tau_f}\!\!\!\!d\tau''
\,\lambda^\Psi_D(y,\tau'') \right]\rule{0pt}{16pt}\!\right\}\!.
\nonumber
\eea
Again, the solution found holds for the different charmonia $J/\psi$, $\psi'$
and $\chi_c$.
The initial condition $dN_\Psi^0/dy$ is given by eq.~(\ref{initialpsi})
and the rates $\lambda^\Psi_D$ and $\lambda^\Psi_F$ are given respectively by 
eqs.~(\ref{rated}) and (\ref{ratef}). Note that they are implicitly 
dependent on impact parameter and collision energy.
The structure of this solution is self-evident. The first term describes
the dissociation of $\Psi$s initially produced in the hard collision, with
the usual exponential suppression acting at all times from $\tau_0$ to 
$\tau_f$, while the second term describes formation
of $\Psi$s from $c \bar c$ in the QGP, from the initial time $\tau_0$ up
to an intermediate value $\tau'$ and their subsequent suppression from 
$\tau'$ to $\tau_f$, integrated over all values of $\tau'$.
This last term may become important as soon as the number of charmed quarks 
is large enough. This is expected to be the case in $AB$ reactions as the 
collision energy increases, since the $c$ quark spectrum is proportional
to the mean number of nucleon-nucleon collision $N_{coll} > 1000$ for 
large nuclei, and because the magnitude of the spectrum grows by nearly
two orders of magnitude from, say, SPS to RHIC energy, as shown in 
Fig.~\ref{figure:numcharm}.
\begin{figure}[t]
\begin{center}
\epsfig{file=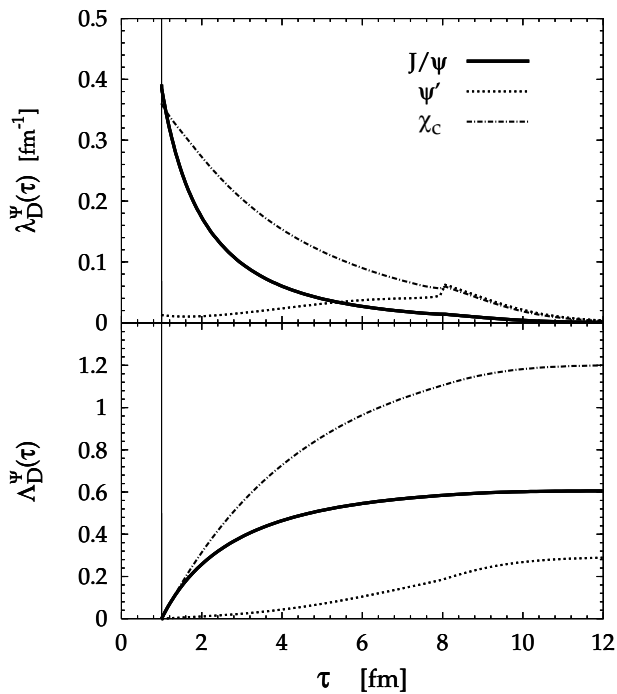,width=8.5cm}
\caption{Top: time evolution of dissociation rates $\lambda^\Psi_D(\tau)$ for
$J/\psi$, $\psi'$ and $\chi_c$, defined in eq.~(\ref{rated}), computed for
central collisions at SPS energy. Bottom: same as above for the integrated
dissociation rates $\Lambda^\Psi_D(\tau)$.}
\label{figure:rate-diss}
\end{center}
\end{figure}

As a final remark we stress that leaving the $y$-integration open
is important since we are interested in computing the value of the final 
$\Psi$ rapidity distribution at mid-rapidity, and not the whole yield,
for which experimental results are available and with which we want to
confront our model.

\subsection{Results}

Using the elements of the calculation as discussed in the previous
sections, we can now compute the time dependence of dissociation and
formation rates. The dissociation rate depends on the fireball
temperature which, in turn, depends on time. How $T$ depends in detail
on $\tau$, was evaluated for different impact parameters and collision
energies in section \ref{sec_fireball}. For central collisions at SPS
energies (for RHIC conditions the result is qualitatively similar),  
the dissociation rates $\lambda^\Psi_D$, defined in
eq.~(\ref{rated}) are plotted against $\tau$ in Fig.~\ref{figure:rate-diss}
(top). For $J/\psi$ it drops fast, initially approximately as $1/\tau$,
for $\chi_c$ the rate drops more slowly, while for $\psi'$ it is maximum at
hadronization. The integrated rates, denoted by $\Lambda^\Psi_D$,
constitute the exponents of the suppression factors in eq.~(\ref{solution}).
They are also plotted in Fig.~\ref{figure:rate-diss} (bottom). 
All curves level off after hadronization in a similar way and
the final value for $\chi_c$ dominates the others. 

\begin{figure}[t]
\begin{center}
\hspace{-4mm}
\epsfig{file=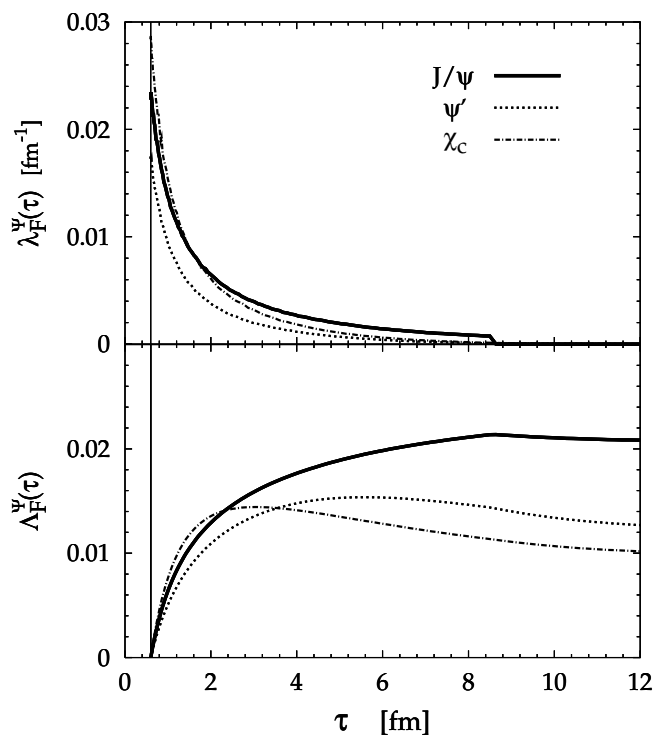,width=9cm}
\caption{Top: time evolution of formation rates $\lambda^\Psi_F(\tau)$ for
$J/\psi$, $\psi'$ and $\chi_c$, defined in eq.~(\ref{ratef}), computed for
central collisions at RHIC energy. Bottom: same as above for the integrated
formation rates $\Lambda^\Psi_D(\tau)$ weighted by the suppression factor as 
in the r.h.s. of eq.~(\ref{solution}).} 
\label{figure:rate-form}
\end{center}
\end{figure}

Analogously, we computed the formation rates $\lambda^\Psi_F$, defined in
eq.~(\ref{ratef}), which depend on time via the factor $1/\tau$ and a
factor $1/R^2(\tau)$, contained in the quantity $K$ given with
eq.~(\ref{kappa}). It turns out that the coalescence mechanism is
ineffective at SPS energies, therefore we plot the formation rate
against time, for central RHIC collisions, in
Fig.~\ref{figure:rate-form} (top). For all charmonia the formation rates drop
faster with increasing $\tau$ than the dissociation rates. They are
negligible well before hadronization occurs. 
The integrated formation rate is folded with the exponential
suppression factor and computed as in the second term on the r.h.s. of
eq.~(\ref{solution}), denoting it by $\Lambda^\Psi_F$. 
The result is plotted also in Fig.~\ref{figure:rate-form} (bottom). 
Obviously all curves level off before hadronization, showing a dominance
of $c \bar c$ coalescence into $J/\psi$ rather than in $\psi'$ or $\chi_c$.
Noteworthy is the tendency of the $\Lambda^\Psi_F$s to decrease once
again at later times, when the $\lambda^\Psi_F$s are by then small and
dissociation tends to take over. This behavior is more pronounced for
$\chi_c$, for which dissociation is still strong in the vicinity of
hadronization.

Dissociation and formation rates are then combined together according
to eq.~(\ref{solution}) to give the observed $\Psi$ rapidity
distribution. As the fireball evolution is constructed for different
impact parameters and collision energies, we compute the corresponding
rapidity distribution at $y = 0$. 

Since we are interested for the final $J/\psi$ yield, we need to take into
account decays into the charmonium ground state of $\psi'$ and $\chi_c$.
Experimentally is has been observed that measured $J/\psi$s come in a fraction
of $60\,\%$ from direct production, while a fraction of $10\,\%$ comes
from $\psi'$ decays and a fraction of $30\,\%$ is from $\chi_c$. Denoting
these fractions as $W_{J/\psi} = 0.6$, $w_{\psi'} = 0.1$ and $w_{\chi_c} =
0.3$, we can write the final $J/\psi$ rapidity distribution as
\be
\frac{dN^{fin}_{J/\psi}}{dy} = \sum_\Psi\ 
W_\Psi\ \frac{dN^{f}_{\Psi}}{dy}\,,
\label{weighted}
\ee
where the sum extends over $\Psi = J/\psi,\psi',\chi_c$. Approximating the
weighted exponential for suppression of the initially produced $J/\psi$s, 
implicit in eq.~(\ref{weighted}) and appearing in eq.~(\ref{solution}), as 
\bea
\sum_\Psi\ W_\Psi\ \frac{dN_{\Psi}^0}{dy}
\,\exp\left[ -\! \int_{\tau_0}^{\tau_f}\!\!\!d\tau'\,
\lambda^\Psi_D(y,\tau')\right] \hspace{18mm} && \\
\hspace{15mm} \simeq\ \frac{dN_{J/\psi}^0}{dy}
\ \sum_\Psi\ W_\Psi\,\exp\left[ -\! \int_{\tau_0}^{\tau_f}\!\!\!d\tau'\,
\lambda^\Psi_D(y,\tau')\right]\!, && \nonumber
\eea
which is correct to leading order in the expansion of the exponential,
we obtain the final form of the $J/\psi$ rapidity distribution which can be
compared with experimental data. The required expression is
\bea
\frac{dN^{fin}_{J/\psi}}{dy} \!\!&=& \!\!
\left\{\rule{0pt}{16pt} \frac{dN_{J/\psi}^0}{dy}
\, \sum_\Psi\, W_\Psi\,\exp\left[ -\! \int_{\tau_0}^{\tau_f}\!\!\!d\tau'\,
\lambda^\Psi_D(y,\tau')\right] \right. \label{finalsolution} \\
&& \left. \hspace{-15mm} +\  \sum_\Psi\,W_\Psi
\,\int_{\tau_0}^{\tau_f}\!\!\!d\tau'\, 
\lambda^\Psi_F(y,\tau')\, \exp\left[ -\! \int_{\tau'}^{\tau_f}\!\!\!\!d\tau''
\,\lambda^\Psi_D(y,\tau'') \right]\rule{0pt}{16pt}\!\!\right\}\!.
\nonumber
\eea

\begin{figure}[t]
\begin{center}
\epsfig{file=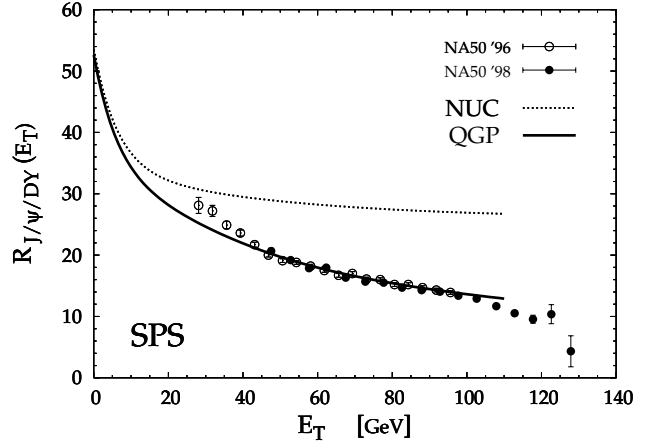,width=6cm,angle=-90}
\caption{Result at SPS energy for the $J/\psi/DY$ ratio as function of
  the transverse energy. The dotted curve, labelled ``$NUC$'', includes only 
  nuclear effects, while the solid line, labelled ``$QGP$'', is the complete 
  result including dissociation by collisions with gluonic quasi-particles.}
\label{figure:result-sps}
\end{center}
\end{figure}

We now look at the case of $Pb\!+\!Pb$ collision at $\sqrt{s} = 17.4$ 
GeV and compute the $J/\psi$ spectrum as function of the impact
parameter $b$. Then we construct the $J/\psi/DY$ ratio 
\bea
R_{J/\psi/DY}(b) 
\!&=&\! B_{\mu^+ \mu^-} \frac{dN_{J/\psi}(b)/dy}{dN_{DY}(b)/dy} \\
&&  \hspace{-2.2cm} = B_{\mu^+ \mu^-} 
\frac{dN_{J/\psi}(b)/dy}{dN_{DY}^{pp}/dy\ N_{coll}(b)}
= N_0\ \frac{dN_{J/\psi}(b)/dy}{dN_\Psi^{pp}/dy\ N_{coll}(b)}, \nonumber 
\eea
where we assume that the Drell-Yan spectrum in $Pb\!+\!Pb$ collisions scales
with the number of collisions $N_{coll}$. The coefficient $B_{\mu^+ \mu^-}$
is the branching of $J/\psi$ into $\mu^+ \mu^-$ pairs. The overall 
normalization is fixed at 
$N_0 = 53.5$ to match, at large values of $b$, the experimental value in 
$pp$ collisions. The $J/\psi/DY$ ratio is then plotted in  
Fig.~\ref{figure:result-sps} (solid line) as function of the mean 
transverse energy
\be
E_T(b) = \epsilon_T\, N_p(b)\,. 
\ee
The quantity $\epsilon_T = 0.274$ is the amount of produced transverse 
energy per participant. As reference, we also plot the curve obtained by 
considering only nuclear effects (dotted line) and 
neglecting the contribution of the produced medium. 
The agreement with data is quite remarkable, in particular the slope of the
curve, considering that fireball 
parameters have not at all been tuned to this particular observable.
The result as such deserves some detailed comments.
First of all, for simplicity we do not perform the usual convolution with 
the  $E_T-b$ correlation function \cite{KLNS96}, therefore it is
obvious that the curves end at $E_T \simeq 110$ GeV, which corresponds to 
the mean transverse energy at $b=0$. To go beyond this point and be in 
agreement with the data it is  necessary to include effects of fluctuations, 
which can be treated in a straightforward manner \cite{fluct}. Second, we 
find that the formation mechanism of $J/\psi$ by $c \bar c$ coalescence is 
totally negligible at this energy, while suppression is caused exclusively by 
collisions with gluonic quasi-particles.
Third, hadronic dissociation of $J/\psi$ seems to be ruled
out, not because cross sections are small, rather because the hadronic
number density is more than an order of magnitude lower than the
partonic one, and only a fraction of a fm$^{-3}$ already at
hadronization.  All this was clearly shown in
Fig.~\ref{figure:density}.  This last statement relies on the fact
that the time evolution of the fireball is not constructed {\it ad
hoc} in order to obtain the desired result, but a priori and with
independent experimental information (hadronic and low-mass dilepton
spectra). Last, the low $E_T$ region is over-estimated in our
result. This is mainly due to the too simple scaling with the number
of participants, neglecting the fact that for central collisions the
amount of produced transverse energy receives a contribution from the
mean number of collisions. In any case, the fireball description is
not expected to be valid at large impact parameters.

\begin{figure}[t]
\begin{center}
\epsfig{file=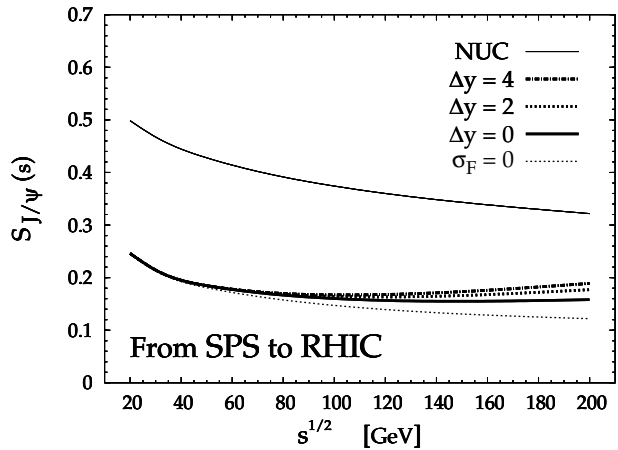,width=6cm,angle=-90}
\caption{Suppression factor $S_{J/\psi}$ as function of the collision
  energy. The thin solid line labelled ``$NUC$'' is computed with nuclear
  effects alone. The thin dotted line labelled ``$\sigma_F\!=\!0$'' is the 
  result with only $J/\psi$ dissociation. The thick solid line labelled 
  ``$\Delta y\!=\!0$'' is the full calculation.
  The thick dotted and dot-dashed lines, labelled ``$\Delta y\!=\!2$'' and
  ``$\Delta y\!=\!4$'' respectively, are obtained by relaxing the 
  assumption of complete $y_c - \eta$ correlation, as explained in the
  text.}
\label{figure:result-rs}
\end{center}
\end{figure}

As we now increase the collision energy, staying at impact parameter $b=0$, 
we observe an interesting feature. Since the amount of charmed quarks
grows substantially with increasing energy, as was indicated  in
Fig.~\ref{figure:numcharm}, we expect that the formation of $J/\psi$s
via  $c \bar c$ coalescence can become important and at least comparable
to the primordial one. To study this we construct the suppression function
\be
S_{J/\psi}(s) = \frac{dN_{J/\psi}(s,b=0)/dy}{dN_{J/\psi}^{pp}(s)/dy
\ N_{coll}(b=0)} 
\ee
and look at its dependence on $s$. Note that in absence of nuclear or medium
effects one should have $S_{J/\psi} \equiv 1$. Indications of the onset of 
the formation mechanism clearly appear from our calculations, which we plot
in Fig.~\ref{figure:result-rs}.
The thin solid line labelled ``$NUC$'' represents the suppression function 
when only nuclear effects are included. According to eq.~(\ref{abscross}) the
effective absorption cross section is 
$\sigma_\Psi^{NUC}(\sqrt{s} = 200\ {\rm GeV})$ = 8.2 mb. The thin dotted line
labelled ``$\sigma_F = 0$'' is the result with only $\Psi$ dissociation,
without the formation mechanism. The thick solid line labelled 
``$\Delta y = 0$'' is the full calculation. One clearly observes, especially 
at the full RHIC energy, a substantial contribution of about 30$\,$\% of
the total from the formation mechanism. The thick dotted and dashed lines,
labelled ``$\Delta y = 2$'' and ``$\Delta y = 4$'' are obtained by relaxing
the assumption of complete $y_c - \eta$ correlation realized by the 
$\delta$-function in eq.~(\ref{cdist}). This is done approximately by smearing
the latter with a Gaussian of width $\Delta y$. The values chosen are 
$\Delta y = 0$, the baseline value, and $\Delta y = 2,4$. More details are
given at the end of the Appendix. Smearing the $y_c - \eta$ correlation 
allows the formation mechanism to become more efficient, leading to a $J/\psi$
yield of comparable magnitude with the primordial one.
The net effect is a slowly rising function of $\sqrt{s}$, but in any case 
we still have suppression, even at RHIC energy. 

\begin{figure}[t]
\begin{center}
\epsfig{file=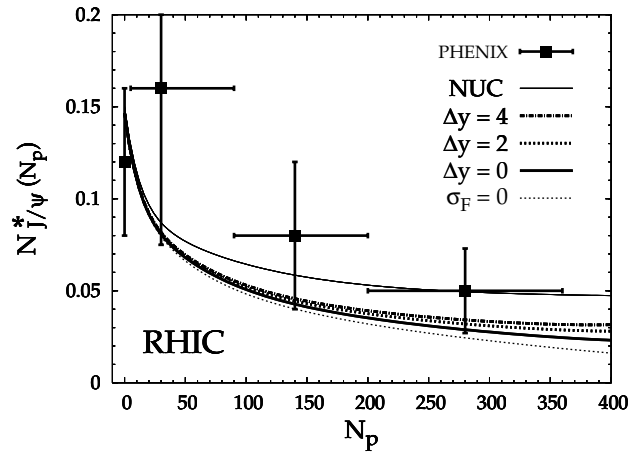,width=6cm,angle=-90}
\caption{Result at RHIC energy for the $J/\psi$ yield, scaled by the number 
of binary collisions, as function of the number of participants. Different
line types for the curves and their labels are described in 
Fig.~\ref{figure:result-rs}. Filled squares are preliminary PHENIX data.}
\label{figure:result-rhic}
\end{center}
\end{figure}

We now proceed to examine central collisions at RHIC energy.
In order to compare with preliminary data from the 
PHENIX experiment \cite{PHENIX}, obtained at mid-rapidity from $\Psi$
decays into $e^+ e^-$ pairs, we construct the quantity
\be
N^{\displaystyle *}_{J/\psi}(b) = B_{e^+ e^-} \frac{dN_{J/\psi}(b)/dy}{N_{coll}(b)} 
\ee
and plot it in Fig.~\ref{figure:result-rhic} as function of the number of 
participants. The quantity $B_{e^+ e^-}$ is the branching of $J/\psi$ into
$e^+ e^-$ pairs, while the overall normalization is fixed at
$N^{\displaystyle *}_{J/\psi}(N_p = 0) = 0.15$. We label all curves 
consistently
with Fig.~\ref{figure:result-rs}. Although a comparison with the data is,
at present, premature, we see that our results lie within experimental
errors. Again, the
contribution from $c \bar c$ coalescence into $J/\psi$ is significant, although
not dramatic. All the obtained curves are monotonically decreasing and we
do not find any inversion of this tendency at any large value of $N_p$.
In other words, we do not find a net enhancement of
$J/\psi$, but this result needs to be confirmed by more accurate
calculations. The reasons are twofold: first, although we have a rough
picture of
what medium to expect at RHIC, the corresponding fireball is presently not 
yet determined at the same level of quality as with existing SPS data. Second,
shadowing effects for $J/\psi$ production have been taken into account
only very crudely. For open
charm production also recent calculations \cite{KT02} show a non
negligible amount of shadowing. Indeed, the expectation at RHIC energy
is a reduction of $dN^c_{AB}/dy$ by about 30$\%$ at mid-rapidity. 
The resulting reduction factor of 0.7 enters squared in the 
formation rate given on the right of eq.~(\ref{ratef}), therefore reducing 
by about half the enhancement effect.

\subsection{Remarks on the QGP at SPS}

Consider once again the solution of the kinetic equation given as
eq.~(\ref{solution}) and neglect the formation term. Taking the time 
evolution of the medium density in the simple form $n(\tau) = n_0
\, \tau_0/\tau$ and assuming that the average dissociation cross
section is time independent (denoted by $\sigma_D$), we 
obtain the simple analytical solution 
\be
\frac{dN_{J/\psi}^f}{dy} = \frac{dN_{J/\psi}^0}{dy}
\, \exp\left[\rule{0pt}{10pt} - \sigma_D\, n_0\,\tau_0\, \log(\,n_0\,/\,n_f\,)\right]\,,
\label{exactsolution}
\ee
where $n_f = n(\tau_f)$. We examine the exponential suppression factor for 
parameter values corresponding to the situation in central collisions at
the SPS. We assume that the produced QGP is thermalized so that the gluon 
density is given, in the ideal gas limit, by
\be
n(T) = \frac{g\, \zeta(3)}{\pi^2}\ T^3\,.
\label{thermglue}
\ee
The final density
is fixed by eq.~(\ref{thermglue}) at the critical temperature $T_c = 170$ MeV,
yielding $n_f = 1.25$ fm$^{-3}$. Taking $\sigma_D = 1$ mb and $\tau_0 = 1$ fm,
consistently with what has been computed earlier in the previous  
sections, one needs $n_0 = 5$ fm$^{-3}$ in order to have a suppression
factor of about 0.5, which is precisely the amount of anomalous suppression 
required by the NA50 data at large $E_T$. Note that the initial density 
required in this scenario corresponds to $T_0 = 270$ MeV.

We now compare the above estimate with the results given by a popular model
\cite{ACF}, describing $J/\psi$ suppression by means of collisions with
comovers, usually considered to belong to the class of ``hadronic models''.
Its essential element is the suppression function
\be
S(\vec b,\vec r) = \exp\left[- \sigma_{co}\, N_y^{co}(\vec b,\vec r)
                     \, \log(\,N_y^{co}(\vec b,\vec r)/N_f\,)\right],
\ee
where $\sigma_{co} = 1$ mb represents the interaction with comovers, 
$N_y^{co}(\vec b,\vec s)$ is the 2-dimensional rapidity density of 
hadrons as function of the transverse coordinate $\vec r$ for a given
impact parameter $\vec b$, computed in the Dual Parton Model \cite{CSTT94}, 
and $N_f = 1.15$ fm$^{-2}$.
Averaging $S$ by means of the ``production probability'' $T_AT_B$ and 
neglecting nuclear effects, we obtain
\be
\langle\,S\,\rangle(\vec b) 
\ \approx\ \exp\left[\rule{0pt}{10pt} - \sigma_{co}\, N_y(b)
\, \log(\,N_y(b)/N_f\,)\right],
\ee
where the average rapidity density of comovers is computed as
\be
N_y(b) = \frac{\displaystyle \int d^2 r
\ T_A(r)\,T_B(\tilde r)\ N_y^{co}(\vec b,\vec r)}
{\displaystyle \int\!\! d^2 r\ T_A(r)\,T_B(\tilde r)}\,.
\label{avgcom}
\ee
If we now identify  
\be 
\sigma_D = \sigma_{co}\ \ ,\ \ \ 
n_0\,\tau_0 = N_y(b=0)\ \ \ \mbox{and}
\ \ \ n_f\,\tau_0 = N_f\,,
\ee
we immediately recover the suppression factor in eq.~(\ref{exactsolution}).
Computing numerically
eq.~(\ref{avgcom}) at $b=0$, we obtain $N_y = 6$ fm$^{-2}$, which should be
compared with the product $n_0\,\tau_0 = 5$ fm$^{-2}$. 
The numerical agreement is quite striking.
To show that this is not a mere coincidence, we can compare the final
values for the densities. Numerically this means to confront $N_f = 1.15$ 
fm$^{-2}$ with $n_f\,\tau_0 = 1.25$ fm$^{-2}$ determined above: again, a 
remarkable agreement. It is clear that the initial density of
``hadrons'' as computed 
with the Dual Parton Model is far too high for the medium to be ``hadronic'',
and the particle density $N_f$ is not to be understood as the freeze-out
value but rather to be taken at hadronization. It is therefore
suggestive to interpret the comover model as actually describing
anomalous $J/\psi$ suppression by the QGP.

\section{Conclusions}

We have calculated $J/\psi$ production from an expanding fireball created in
relativistic heavy-ion collisions over a wide range of centralities and
beam energies, from SPS at $\sqrt{s} =$ 17.3 GeV ($E_{lab} =$ 158 GeV) 
to RHIC at $\sqrt{s} =$ 200 GeV. In this summary we review our basic
assumptions put them into perspective. 

The produced medium was described assuming QGP formation and thermalization. 
A phenomenological quasi-particle model for quarks and gluons, 
in accordance with lattice QCD thermodynamics, was applied to model the
partonic phase. This provided a realistic EoS which we then used to
drive the expansion dynamics of the medium by
means of a fireball model, characterized by time dependent temperature
and volume. This construction
had already been proven successful in describing low mass dilepton
spectra at SPS. We want to stress again that the setup of this fireball
(initial and freeze-out temperature, density etc.) had been fixed by
independent observables and ensures consistency with a multitude of
hadronic measurements. In contrast to previous approaches, we have
therefore largely eliminated the medium evolution as an adjustable parameter 
in the $J/\psi$ description. Nevertheless, this framework is still
simplified and is probably insufficient to account for phase-space
correlations of the medium, which have been averaged out. In a
future attempt to study rapidity and $p_\perp$-dependent properties of
$J/\psi$, this is a problem to be taken into account.

We then set up a kinetic equation with a collision term incorporating both gain
and loss terms. These stand for $\Psi$ formation due to
coalescence of $c \bar c$ quark pairs and $\Psi$
dissociation due to collisions with gluon quasi-particles, respectively.
The elementary
process $\Psi g \rightarrow c \bar c$ was modeled by a simplified
dissociation cross section, approximating $\Psi$ as a Coulomb bound
state. The corresponding back reaction $c \bar c \rightarrow \Psi g$
could then be obtained by detailed balance. Here a few important
remarks are necessary. First of all, a crucial ingredient for the
construction of the solution of the kinetic equation is a reliable
initial condition. In the present work we have simplified this aspect 
in order to
focus on the QGP effects, but it is possible to improve on this point,
especially in what concerns the description of nuclear effects, both
for open and for hidden charm. New data collected at RHIC in recent
deuteron+$Au$ collisions will certainly help constraining the model
dependence of such calculations. Second, the description of medium effects
on $\Psi$ was developed neglecting screening of the $c \bar c$
potential. This is an assumption which should be checked by a
comprehensive study of the $\Psi$ self energy in a QGP, including, at
the same time, screening and collision processes. Third, the use of the 
Coulomb part of the potential to describe $c \bar c$ bound states is only
a rough approximation. In fact, cross sections involving $\psi'$ and
$\chi_c$ had to be rescaled in order to achieve agreement with the
data. A proper description of bound states with a realistic potential,
including a confining and possibly screened part, is then necessary.
Despite such simplifications, the description as a whole seems to be
coherent and provides a framework for future developments.

With proper averaging of the kinetic equation over the spatial extent
of the fireball, we then found a simple solution,
which allowed direct comparison with experiment. At SPS energy 
we were able to describe the suppression effect in the data, 
without the need to invoke hadronic
comovers. These results support the hypothesis that the QGP is actually
produced, at a transient stage, in $Pb\!+\!Pb$ collisions at 
$\sqrt{s}=$ 17.3 GeV.
It is important to note that, within our outlined approach, a purely hadronic
framework would not be successful in describing existing data. Moreover,
since the hadronic phase exists only at moderately low particle densities,
it has no bearing on $\Psi$ evolution. We also considered extrapolations
up to RHIC energies where, despite the more extreme conditions as
compared to SPS, a sizable fraction of primordial $J/\psi$s still survives. 
Although a clear trend towards more copious $J/\psi$ production in by 
$c \bar c$
coalescence was found, no net $J/\psi$ enhancement was present in the end.  
This result needs to be confirmed by a dedicated study of space-time
and momentum $\eta-y_c$ rapidity correlations for charm quarks, since
they have a non-negligible effect on the final results.

\section*{Acknowledgements}

We are grateful to Peter Braun-Munzinger, Elena Ferreiro, Pol-Bernard
Gossiaux, J\"org H\"ufner and Helmut Satz
for stimulating discussions and in particular to David Blaschke for numerous
exchanges of opinions during the preparation of this paper. 
This work was supported in part by BMBF and GSI.

\begin{widetext}

\appendix*
\section{Evaluation of the coalescence rate}

We evaluate here the coalescence rate. From the definition of the formation
term in eq.~(\ref{collf}), substituting in it eqs.~(\ref{ph3}), (\ref{crossw})
and (\ref{cdist}) we obtain
\bea
C_F(p,x) &=& \frac{2 \pi^2}{\tau^2}
\int\! \frac{d^3 k}{E_k}\,\frac{d^3 q_1}{E_1}
\,\frac{d^3 q_2}{E_2}\ \delta^4(p+k-q_1-q_2)
\ S_c(y_1,q_\perp^1)\,S_c(y_2,q_\perp^2)
\ \delta(y_1-\eta)\,\delta(y_2-\eta) \\
&& \hspace{-1cm}
\times\ \frac{s\, F_{c \bar c}(s)}{m_\perp^1 m_\perp^2\, F_{\Psi g}(s)}
\ \sigma_F(s) 
\ \frac{1}{\left[\rule{0pt}{9pt} \pi R^2(\tau) \right]^2}
\,\theta\!\left(\rule{0pt}{9pt} R(\tau) - 
|\vec r_\perp - (\vec q_\perp^1 / m_\perp^1)\,\tau|\right)\,
\theta\!\left(\rule{0pt}{9pt} R(\tau) - 
|\vec r_\perp - (\vec q_\perp^2 / m_\perp^2)\,\tau|\right). \nonumber
\eea
Integrating over all phase space variables, except for $y$, we define the 
formation rate as
\bea
\lambda_F(\tau) &=& \left. \frac{1}{(2 \pi)^3} 
\int\! d^2 p_\perp \!\int\! d^2 r_\perp d\eta\ \tau \ C_F(p,x) 
\ \right|_{\, y \, =\,  0}\nonumber \\
&=&\frac{1}{4 \pi \tau}
\int\! d^2 k_\perp\,d^2  q^1_\perp\,d^2  q^2_\perp
\int\!dy_k\,dy_1\,dy_2\ \delta(E_\Psi+E_k-E_1-E_2)
\, \delta(p_z+k_z-q_z^1-q_z^2)\, \delta(y_1-y_2) \nonumber \\
&& \hspace{2.5cm} \times\ \frac{s\,(s - 4m_c^2)}{m_\perp^1 m_\perp^2
\,(s - m_\Psi^2)}
\ \sigma_F(s) \ S_c(y_1,q_\perp^1)\,S_c(y_2,q_\perp^2) \nonumber \\
&& \hspace{2.5cm} \times\ 
\int\! d^2 r_\perp\ \frac{1}{\left[\rule{0pt}{9pt} 
\pi R^2(\tau) \right]^2}\,\theta\!\left(\rule{0pt}{9pt} R(\tau) - 
|\vec r_\perp - \vec \xi/2|\right)\,\theta\!\left(\rule{0pt}{9pt} R(\tau) - 
|\vec r_\perp + \vec \xi/2|\right), 
\label{appform1}
\eea
where $\vec k_\perp = \vec q^1_\perp + \vec q^2_\perp - \vec p_\perp$
and $\vec \xi = (\vec q_\perp^1 / m_\perp^1 - \vec q_\perp^2 / m_\perp^2)
\,\tau$. The integral over rapidities can be manipulated to give
\bea
\int\!dy_k\,dy_1\,dy_2\ \delta(m^\Psi_\perp+k \cosh y_k-m_\perp^1 \cosh y_1
-m_\perp^2 \cosh y_2)\nonumber \\
\times \,\delta(k \sinh y_k-m_\perp^1 \sinh y_1
-m_\perp^2 \sinh y_2)\, \delta(y_1-y_2) && \nonumber \\
&&\hspace{-10cm}  
=\ \frac{4}{\sqrt{(\mu_N^2)^2 - (\mu_D^2)^2}} \int\!dy_1\,dy_2
\ \delta(y_1-\bar y_c) \,\delta(y_2-\bar y_c)\,\theta(\mu_N^2-\mu_D^2),
\label{appform2}
\eea
where $\mu_N^2 = (m_\perp^\Psi)^2 + (m_\perp^1+m_\perp^2)^2 - k_\perp^2$,
$\mu_D^2 = 2 m_\perp^\Psi(m_\perp^1+m_\perp^2)$ and
$\bar y_c = \log\left(\mu_N^2/\mu_D^2 + \sqrt{\rule{0pt}{9pt} 
(\mu_N^2/\mu_D^2)^2 -1}\right)$. The integral over the transverse position
variable can be computed analytically and it turns out to depend only on the
variable $x = \xi/2R(\tau)$. The result of the integration is
\be
Q(x) = \frac{1}{\pi R^2(\tau)}\,\left[ 1 - \frac{2}{\pi} 
\left( x \sqrt{1 - x^2} + \arcsin x \right) \right]\,\theta(1-x).
\ee
In this way the formation rate can be brought to the form
\bea
\lambda_F(\tau) &=& \frac{1}{\tau}
\int\! d^2 k_\perp\,d^2 q^1_\perp\,d^2 q^2_\perp
\ \frac{s\,(s - 4m_c^2)}{\pi\,m_\perp^1 m_\perp^2\,(s - m_\Psi^2)}
\ \sigma_F(s)\ S_c(\bar y_c,q_\perp^1)\,S_c(\bar y_c,q_\perp^2)
\\
&&\times \ \frac{Q(x)}{\sqrt{(\mu_N^2)^2 -
(\mu_D^2)^2}}\,\theta(\mu_N^2-\mu_D^2). \nonumber
\eea
Defining the quantity
\be
K = \frac{s\,(s - 4m_c^2)}{\pi\,m_\perp^1 m_\perp^2\,(s - m_\Psi^2)}
\ \frac{ Q(x)}{\sqrt{(\mu_N^2)^2 - (\mu_D^2)^2}}
\,\theta(\mu_N^2-\mu_D^2),
\label{kappa}
\ee
we finally arrive at the expression given in eq.~(\ref{ratef}).

If we now relax the assumption that the $c$ quark distribution contains the 
factor $\delta(y_c - \eta)$, replacing it with a Gaussian
$g(y_c - \eta) = (\pi \Delta y^2)^{-1/2} \exp[-(y_c - \eta)^2/\Delta y^2]$,
the result of the $\eta$-integral, which is one of those to be done to obtain
eq.~(\ref{appform1}), is
\be
\int\!d\eta\ g(y_1 - \eta)\,g(y_2 - \eta) = (2 \pi \Delta y^2)^{-1/2}\,
\exp\left[-(y_c - \eta)^2/2\Delta y^2\right].
\ee
Defining $\rho = y_1 - y_2$ and $y_c = (y_1 + y_2)/2$ and expanding the 
l.h.s. of eq.~(\ref{appform2}) to order $\rho^2$, we approximate the various
terms typically as
\bea
m_\perp^1 \cosh y_1 + m_\perp^2 \cosh y_2 &\simeq& (m_\perp^1 + m_\perp^2)
\,\cosh y_c\ (1 + \rho^2 / 8)
\ \ \ {\rm and}\ \ \ \\
m_\perp^1 \sinh y_1 + m_\perp^2 \sinh y_2 &\simeq& (m_\perp^1 + m_\perp^2)
\,\sinh y_c\ (1 + \rho^2 / 8).
\eea
We have assumed that $m_\perp^1 - m_\perp^2 \simeq 0$ for the values of
$q_\perp^1$ and $q_\perp^2$ which dominate the final phase-space integration.
Substituting $\rho^2$ with the mean value $\Delta y^2$, the final result is
reduced to the substitution
\be
m_\perp^1 + m_\perp^2 \longrightarrow (m_\perp^1 + m_\perp^2)
\ (1 + \Delta y^2 / 8)
\ee
in $\mu_N^2$. This amounts to open up phase-space for the transverse
integrations since there are more values of transverse momenta which 
satisfy the constraint $\theta(\mu_N^2-\mu_D^2)$.

\end{widetext}

\end{document}